\documentclass[11pt]{article}
\usepackage[utf8]{inputenc}
\usepackage[authoryear]{natbib}
\usepackage[a4paper]{geometry}
\usepackage[title]{appendix}
\usepackage{xcolor}
\usepackage{setspace}
\usepackage{algorithm}
\usepackage{algpseudocode}
\usepackage{tikz}
\usetikzlibrary{positioning, fit, backgrounds}
\algrenewcommand\algorithmicrequire{\textbf{Input:}}
\algrenewcommand\algorithmicensure{\textbf{Output:}}

\usepackage{amsmath}
\usepackage{amssymb}

\usepackage{amsfonts}
\usepackage{bbm}
\usepackage{graphicx}
\usepackage{booktabs, multirow}
\usepackage{makecell}
\usepackage{amsthm}
\usepackage{url}
\usepackage{comment}
\usepackage{subcaption}
\usepackage{float}
\usepackage{cleveref}
\usepackage{mathtools}
\usepackage{placeins}
\mathtoolsset{showonlyrefs=true}

\numberwithin{equation}{section}

\theoremstyle{plain}
\newtheorem{lemma}{Lemma}

\newtheorem{theorem}{Theorem}
\newtheorem{corollary}{Corollary}
\newtheorem{proposition}{Proposition}
\newtheorem{assumption}{Assumption}

\newcommand{\ELBO}{\mathrm{ELBO}}

\newcommand{\E}{\mathbb E}
\newcommand{\vgamma}{\boldsymbol{\gamma}}
\newcommand{\vlambda}{\boldsymbol{\lambda}}

\title{Bayesian variable selection via collapsed importance-weighted variational inference over discrete model spaces}
\author{Mai Mai\thanks{The University of Sydney Business School, Sydney, NSW, Australia.} \and Matias Quiroz\thanks{School of Mathematical and Physical Sciences, University of Technology Sydney, NSW, Australia.}$\;\,$\thanks{Human Technology Institute, University of Technology Sydney.} \and Minh-Ngoc Tran$^*$ }
\date{}

\begin{document}

\maketitle

\begin{abstract}
Bayesian variable selection under spike-and-slab priors provides interpretable posterior inclusion probabilities for individual predictors, but exploring large discrete model spaces using conventional Markov chain Monte Carlo methods can be computationally challenging. With a particular focus on posterior inclusion probability estimation, we develop a collapsed importance-weighted variational framework for Gaussian linear regression. The framework analytically integrates out the remaining model parameters and directly approximates the posterior over binary inclusion indicators using a factorised Bernoulli distribution. To improve posterior inclusion probability estimation while retaining the computational simplicity of the factorised variational family, we replace the conventional evidence lower bound with a tighter variational R\'enyi importance-weighted lower bound. We establish a finite-sample upper bound on posterior inclusion probability approximation errors and show that, under suitable conditions, the optimised variational inclusion probabilities converge to their exact posterior counterparts as the number of importance samples increases and the R\'enyi parameter decreases appropriately. We further establish that analytical marginalisation tightens the variational objective and the corresponding inclusion probability error bound, while reducing importance-weight variance and improving the asymptotic efficiency of importance-sampling post-processing. To address the challenges of estimating gradients with respect to the variational parameters when the inclusion indicators are discrete, we leverage recent advances in variance-reduced score-function estimators, together with an annealing strategy to stabilise optimisation. Simulation experiments and empirical applications demonstrate improvements in posterior inclusion probability estimation relative to standard variational inference and two existing variational approaches.
\end{abstract}

\section{Introduction}\label{sec:Introduction}

Bayesian variable selection via spike-and-slab priors (see \citeauthor{ohara2007review},~\citeyear{ohara2007review} for an overview) provides a principled framework for quantifying uncertainty over the inherently discrete space of candidate models through latent binary inclusion indicators. This yields interpretable posterior inclusion probabilities and coherent Bayesian model averaging. However, posterior inference in high dimensions remains computationally challenging due to both the combinatorial size of the induced model space and the difficulty of efficiently exploring its discrete posterior landscape. Consequently, practitioners often favour continuous shrinkage priors, such as horseshoe-type priors \citep{carvalho2010horseshoe, Bhadra2017horseshoeplus}, which offer a computationally attractive alternative with strong predictive performance. However, these approaches typically do not directly target posterior inference over discrete model configurations. In this paper, we develop an importance-weighted variational framework for Bayesian variable selection in Gaussian linear regression that operates directly over the binary inclusion indicators after analytically integrating out the remaining model parameters, with a particular emphasis on estimating posterior inclusion probabilities.

Posterior inference under exact spike-and-slab priors has traditionally relied on Markov chain Monte Carlo methods, with foundational contributions in linear regression \citep{mitchell1988bayesian} and nonparametric regression \citep{smith1996nonparameteric}. Although these methods avoid explicit enumeration of all possible models, efficiently exploring the discrete posterior distribution becomes increasingly difficult as the number of candidate predictors grows. Variational inference offers a computationally attractive alternative by replacing posterior simulation with optimisation over a tractable approximating distribution; see \citet{blei2017variational} for an overview and \citet{tran2021practical} for a practical tutorial. Existing variational approaches to Bayesian variable selection include \citet{carbonetto2012scalable}, who consider both linear and logistic regression, \citet{ormerod2017variational} and \citet{you2023approximate} for linear regression, \citet{zhang2019variational} for logistic regression, and \citet{bianco2026scalable} for time-varying parameter models. In particular, \citet{you2023approximate} develop an approximate collapsed variational Bayes approach that, like ours, integrates out model parameters before approximating the posterior over inclusion indicators. A key motivation for our work comes from the findings of \citet{carbonetto2012scalable}. They report inaccurate variational estimates of posterior inclusion probabilities under correlated predictors, attributing this limitation to their factorised variational approximation, and identify relaxing the independence assumption as a natural extension of their work. Such an extension could involve more flexible variational approximations that capture posterior dependence, potentially through techniques such as normalising flows \citep{rezende2015normalising} or Gaussian copulas \citep{han2016variational}. However, increased flexibility may come at the expense of computational scalability, particularly in high-dimensional discrete model spaces.

The limitation identified by \citet{carbonetto2012scalable} and the potential scalability challenges of the proposed extensions raise a natural question: Can posterior inclusion probability (PIP) estimation be improved without relaxing the independence assumption in the variational family? We investigate a complementary strategy that retains a simple factorised approximation, consistent with the emphasis on explicit and compact variational representations \citep{zenn2024differentiable}, while seeking to improve PIP estimation through the optimisation objective itself. Specifically, we adopt the variational R\'enyi importance-weighted (VR-IWAE) objective \citep{daudel2023alpha}, which provides a tighter lower bound on the log marginal likelihood than the standard evidence lower bound. We establish a finite-sample upper bound on PIP approximation errors and show that, under suitable conditions, the optimised variational inclusion probabilities converge to their exact posterior counterparts as the number of importance samples increases and the R\'enyi tempering parameter decreases appropriately. This provides an important complement to the findings of \citet{carbonetto2012scalable}, who report deteriorating PIP accuracy under increasing predictor correlation when optimising the standard evidence lower bound. Our result implies that such deterioration need not persist when replacing the standard objective with VR-IWAE: accurate PIP estimation remains possible asymptotically, in the regime described above, even in the presence of posterior dependence among inclusion indicators. This provides theoretical support for retaining a scalable factorised variational family. We further provide a theoretical justification for our collapsed approach by showing that analytical marginalisation yields an importance-weighted variational bound at least as tight as its uncollapsed counterpart. We also establish that collapsing reduces importance-weight variance and improves the asymptotic efficiency of subsequent importance-sampling post-processing.

Our paper makes four main contributions. First, we develop a collapsed importance-weighted variational framework for Bayesian variable selection that operates directly over discrete inclusion indicators, supported by theoretical results on PIP approximation accuracy. Second, we establish theoretical advantages of analytical marginalisation for both variational inference and importance-sampling post-processing. Third, we provide a practical implementation of importance-weighted variational inference for high-dimensional discrete model spaces. Since standard reparameterisation-based gradient estimators are not directly applicable and naive score-function estimators can become increasingly variable as the number of importance samples grows, we employ the variance-reduced score-function gradient estimators of \citet{daudel2026importance}, together with an annealing strategy to stabilise optimisation. Finally, we combine the resulting variational approximation with importance-sampling post-processing, including adaptations of existing proposal-updating strategies to discrete model spaces. We demonstrate the usefulness of the methodology through simulation studies and empirical applications. In particular, we show improvements in PIP estimation relative to both standard ELBO optimisation, which is a special case of our objective, and the variational approaches in \citet{carbonetto2012scalable} and \citet{you2023approximate}.

The remainder of the paper is organised as follows. Section~\ref{sec:Methodology} presents the methodology and accompanying theoretical results. Section~\ref{sec:linear-regression-experiments} reports experiments assessing different aspects of the methodology, while Section~\ref{sec:Applications} presents the real-data applications. Section~\ref{sec:Conclusions} concludes the paper and outlines future research. Proofs, implementation details and additional empirical results are provided in the supplementary material. Equations, sections and other numbered results in the main text are referenced as (1.1), Section~1, etc., while those in the supplementary material use the prefix S, as in (S1.1) and Section~S1. 
An implementation of the proposed methodology is available on GitHub\footnote{\url{https://github.com/hhmai12/collapsed-bvs-iwvi}}.

\section{Methodology}\label{sec:Methodology}

\subsection{Bayesian variable selection in Gaussian linear regression and the fully collapsed posterior}
\label{sec:collapsed-posterior}

Consider Gaussian linear regression with $p$ candidate predictors. Let $\mathbf y\in\mathbb R^n$ denote the response vector and $\mathbf X\in\mathbb R^{n\times p}$ the design matrix. Bayesian variable selection introduces a binary inclusion vector $\vgamma=(\gamma_1,\ldots,\gamma_p)^\top\in\{0,1\}^p$, where $\gamma_j=1$ if predictor $j$ is included in the regression model and $\gamma_j=0$ otherwise. Each configuration $\vgamma$ corresponds to one of $2^p$ possible regression models.

For a given $\vgamma$, define the active set of predictors
\[
\mathcal A=\{j:\gamma_j=1\},
\qquad
p_{\mathcal A}=|\mathcal A|=\sum_{j=1}^p\gamma_j.
\]
Let $\mathbf X_{\mathcal A}\in\mathbb R^{n\times p_{\mathcal A}}$ denote the submatrix of $\mathbf X$ containing the columns indexed by $\mathcal A$, and let $\boldsymbol{\beta}_{\mathcal A}\in\mathbb R^{p_{\mathcal A}}$ denote the corresponding regression coefficients. We assume that the predictors are centred and standardised, while the response is left uncentred. Conditional on the model configuration, we consider the Gaussian regression model
\begin{equation}
\mathbf y|\beta_0,\boldsymbol{\beta}_{\mathcal A},
\sigma^2,\vgamma
\sim
N\left(
\beta_0\mathbf 1_n+
\mathbf X_{\mathcal A}\boldsymbol{\beta}_{\mathcal A},
\sigma^2\mathbf I_n
\right),
\label{eq:bvs-model}
\end{equation}
where the intercept $\beta_0$ is always included and is not subject to variable selection.

We adopt a spike-and-slab prior in which excluded coefficients are identically zero, while the coefficients of the included predictors follow a Gaussian prior distribution,
\begin{equation}
\boldsymbol{\beta}_{\mathcal A}|\sigma^2,\vgamma
\sim
N\left(\mathbf 0,\sigma^2\mathbf I_{p_{\mathcal A}}\right).
\label{eq:bvs-slab-prior}
\end{equation}
The intercept and residual variance are assigned the improper priors
\[
p(\beta_0)\propto 1,
\qquad
p(\sigma^2)\propto\frac{1}{\sigma^2}.
\]
To account for uncertainty about the degree of sparsity, we specify the hierarchical prior
\begin{equation}
\gamma_j|\pi
\overset{\mathrm{ind}}{\sim}\operatorname{Bernoulli}(\pi),
\quad j=1,\ldots,p,
\qquad
\pi\sim\operatorname{Beta}(a_\pi,b_\pi),
\label{eq:bvs-inclusion-prior}
\end{equation}
where $\pi$ denotes the common prior inclusion probability. Integrating out $\pi$ gives the marginal prior probability of a model configuration,
\begin{equation}
p(\vgamma)
=
\frac{B(a_\pi+p_{\mathcal A},b_\pi+p-p_{\mathcal A})}
{B(a_\pi,b_\pi)},
\label{eq:bvs-model-prior}
\end{equation}
where $B(\cdot,\cdot)$ denotes the beta function.

The main computational challenge lies in inference over the inclusion indicators $\vgamma$, since the posterior distributions of the remaining model parameters can be sampled from standard conditional distributions once $\vgamma$ is given. Moreover, the prior specification above allows the intercept, active regression coefficients, residual variance, and prior inclusion probability to be integrated out analytically. This yields the fully collapsed posterior distribution
\begin{equation}
p(\vgamma|\mathbf y)
\propto
p(\mathbf y|\vgamma)p(\vgamma),
\label{eq:bvs-collapsed-posterior}
\end{equation}
where $p(\vgamma)$ is given in \eqref{eq:bvs-model-prior}.

To evaluate the marginal likelihood $p(\mathbf y|\vgamma)$ in \eqref{eq:bvs-collapsed-posterior}, let $\mathbf y_c=\mathbf y-\bar y\mathbf 1_n$ denote the centred response, where $\bar y=n^{-1}\mathbf 1_n^\top\mathbf y$. Since the predictors are centred, the intercept can be integrated out separately. Define
\begin{align}
\mathbf M_{\mathcal A}
&=
\mathbf X_{\mathcal A}^\top\mathbf X_{\mathcal A}
+\mathbf I_{p_{\mathcal A}},
\qquad
\mathbf b_{\mathcal A}
=
\mathbf X_{\mathcal A}^\top\mathbf y_c,
\label{eq:bvs-collapsed-quantities}\\
R_{\mathcal A}
&=
\mathbf y_c^\top\mathbf y_c
-
\mathbf b_{\mathcal A}^\top
\mathbf M_{\mathcal A}^{-1}
\mathbf b_{\mathcal A}.
\nonumber
\end{align}
Integrating out $\beta_0$, $\boldsymbol{\beta}_{\mathcal A}$, and $\sigma^2$ gives the marginal likelihood
\begin{equation}
p(\mathbf y|\vgamma)
\propto
|\mathbf M_{\mathcal A}|^{-1/2}
R_{\mathcal A}^{-(n-1)/2},
\label{eq:bvs-marginal-likelihood}
\end{equation}
where $\mathcal A$ is the active set determined by $\vgamma$, and the proportionality constant does not depend on the model configuration. For the empty model, we adopt the conventions $|\mathbf M_{\varnothing}|=1$ and $R_{\varnothing}=\mathbf y_c^\top\mathbf y_c$. Although the priors on $\beta_0$ and $\sigma^2$ are improper, their unspecified multiplicative constants are common to all model configurations and therefore cancel when the posterior model probabilities are normalised over the model space.

A primary quantity of interest is the posterior inclusion probability (PIP) of predictor $j$,
\begin{equation}
\Pr(\gamma_j=1|\mathbf y)
=
\sum_{\vgamma_{-j}\in\{0,1\}^{p-1}}
p(\gamma_j=1,\vgamma_{-j}|\mathbf y),
\label{eq:bvs-pip}
\end{equation}
where $\vgamma_{-j}$ denotes the vector of inclusion indicators excluding $\gamma_j$. 
Although the collapsed joint density $p(\mathbf y,\vgamma)$ can be evaluated analytically up to a common multiplicative constant, normalising the posterior in \eqref{eq:bvs-collapsed-posterior} requires summation over $2^p$ model configurations, making direct sampling from the posterior computationally prohibitive as $p$ increases. Markov chain Monte Carlo methods avoid computing the normalising constant, but may struggle to efficiently explore the vast discrete model space. These challenges motivate a scalable approximation to the posterior distribution $p(\vgamma|\mathbf y)$, which we develop in the next section using importance-weighted variational inference.

\subsection{Importance-weighted variational inference for the collapsed posterior}
\label{sec:importance-weighted-vi}

We approximate the fully collapsed posterior $p(\vgamma|\mathbf y)$ derived in Section~\ref{sec:collapsed-posterior} using a product-Bernoulli variational distribution,
\begin{equation}
q_{\vlambda}(\vgamma)
=
\prod_{j=1}^p
\rho_j^{\gamma_j}(1-\rho_j)^{1-\gamma_j},
\qquad
\rho_j=\operatorname{logit}^{-1}(\lambda_j),
\label{eq:variational-distribution}
\end{equation}
where $\vlambda=(\lambda_1,\ldots,\lambda_p)^\top\in\mathbb R^p$ denotes the variational parameters. The probabilities $\rho_j$ provide direct approximations to the posterior inclusion probabilities $\Pr(\gamma_j=1|\mathbf y)$ in \eqref{eq:bvs-pip}. We seek to improve the accuracy of these approximations without increasing the complexity of the variational family. To this end, we consider an importance-weighted variational objective, which we show below provides a direct upper bound on PIP approximation errors.

The standard variational approach maximises the evidence lower bound (ELBO; see for example \citealp{blei2017variational}),
\begin{equation}
\ELBO(\vlambda)
=
\E_{q_{\vlambda}}
\left[
\log
\frac{p(\mathbf y,\vgamma)}
{q_{\vlambda}(\vgamma)}
\right]
=
\log p(\mathbf y)
-
\operatorname{KL}
\left(
q_{\vlambda}\|p(\vgamma|\mathbf y)
\right),
\label{eq:ELBO-def}
\end{equation}
where $\operatorname{KL}$ denotes the Kullback--Leibler divergence. Maximising the ELBO therefore minimises the KL divergence from the variational approximation to the posterior. This criterion tends to favour mode-seeking approximations, concentrating probability mass around particular posterior modes while potentially assigning insufficient probability to alternative model configurations. Such behaviour is undesirable when inference relies on posterior inclusion probabilities and other model-averaged quantities.

Importance-weighted variational inference provides an alternative to the standard ELBO by optimising a tighter lower bound on the log marginal likelihood \citep{burda2015importance}. Define the importance weight
\begin{equation}
w_{\vlambda}(\vgamma;\mathbf y)
=
\frac{p(\mathbf y,\vgamma)}
{q_{\vlambda}(\vgamma)}
=
\frac{p(\mathbf y|\vgamma)p(\vgamma)}
{q_{\vlambda}(\vgamma)}.
\label{eq:importance-weight}
\end{equation}
We adopt the variational R\'enyi importance-weighted (VR-IWAE) objective of \citet{daudel2023alpha}, which extends the variational R\'enyi bound \citep{li2016renyi} by incorporating importance weighting while retaining the parameter $\alpha$ to control the influence of the importance weights. For $N\geq1$ independent draws $\vgamma_1,\ldots,\vgamma_N$ from $q_{\vlambda}$, the objective is
\begin{equation}
\ell_N^{(\alpha)}(\vlambda)
=
\frac{1}{1-\alpha}
\E_{\vgamma_{1:N}\sim q_{\vlambda}}
\left[
\log\left(
\frac{1}{N}
\sum_{i=1}^N
w_{\vlambda}(\vgamma_i;\mathbf y)^{1-\alpha}
\right)
\right],
\qquad
\alpha\in[0,1).
\label{eq:VR-IWAE-bound}
\end{equation}
The ordinary importance-weighted evidence lower bound (IWAE) in \cite{burda2015importance} is recovered when $\alpha=0$, while $N=1$ gives the standard ELBO. For a fixed variational distribution, the bounds satisfy the following ordering \citep[Proposition~1]{daudel2023alpha}:
\begin{equation}
\ELBO(\vlambda)
\leq
\ell_N^{(\alpha)}(\vlambda)
\leq
\ell_{N+1}^{(\alpha)}(\vlambda)
\leq
\mathcal L_{\mathrm R}^{(\alpha)}(\vlambda)
\leq
\log p(\mathbf y),
\label{eq:VR-IWAE-ordering}
\end{equation}
where
\begin{equation}
\mathcal L_{\mathrm R}^{(\alpha)}(\vlambda)
=
\frac{1}{1-\alpha}
\log
\E_{q_{\vlambda}}
\left[
w_{\vlambda}(\vgamma;\mathbf y)^{1-\alpha}
\right]
\label{eq:renyi-bound}
\end{equation}
is the variational R\'enyi bound \citep{li2016renyi}. For fixed $N$ and $q_{\vlambda}$, the VR-IWAE bound also increases as $\alpha$ decreases. These properties motivate using a relatively large $N$ and progressively decreasing $\alpha$ during optimisation. However, when the variational approximation is initially far from the posterior, a few importance weights may dominate for small $\alpha$, leading to unstable stochastic optimisation. We address this through the annealing strategy described in Section~\ref{sec:stochastic-optimisation}.

Unlike the ELBO, maximising the VR-IWAE objective for $N>1$ is not equivalent to minimising the KL divergence. 
Its connection to posterior approximation accuracy can instead be established by relating the gap between the VR-IWAE bound and the log marginal likelihood to the R\'enyi divergence. For $0<\alpha<1$, the variational R\'enyi bound can be expressed as \citep{li2016renyi}
\begin{equation}
\mathcal L_{\mathrm R}^{(\alpha)}(\vlambda)
=
\log p(\mathbf y)
-
\frac{\alpha}{1-\alpha}
D_{1-\alpha}
\left(
p(\vgamma|\mathbf y)\|q_{\vlambda}
\right),
\label{eq:renyi-divergence-identity}
\end{equation}
where $D_{1-\alpha}$ denotes the R\'enyi divergence of order $1-\alpha$. Combining this identity with \eqref{eq:VR-IWAE-ordering} gives
\begin{equation}
\log p(\mathbf y)
-
\ell_N^{(\alpha)}(\vlambda)
\geq
\frac{\alpha}{1-\alpha}
D_{1-\alpha}
\left(
p(\vgamma|\mathbf y)\|q_{\vlambda}
\right).
\label{eq:VR-IWAE-renyi}
\end{equation}
This relationship yields the following bound on the posterior inclusion probability approximation errors.

\begin{lemma}
\label{lem:VR-IWAE-PIP}
For $N\geq1$, $0<\alpha<1$ and $\vlambda\in\mathbb R^p$, the errors in approximating the posterior inclusion probabilities in \eqref{eq:bvs-pip} by $\rho_j$ in \eqref{eq:variational-distribution} satisfy
\begin{equation}
\left|
\Pr(\gamma_j=1|\mathbf y)-\rho_j
\right|
\leq
\sqrt{
\frac{
\log p(\mathbf y)
-
\ell_N^{(\alpha)}(\vlambda)
}{
2\alpha
}
},
\qquad j=1,\ldots,p,
\label{eq:VR-IWAE-PIP-bound}
\end{equation}
where $\ell_N^{(\alpha)}(\vlambda)$ is the VR-IWAE objective defined in \eqref{eq:VR-IWAE-bound}.
\end{lemma}

Lemma~\ref{lem:VR-IWAE-PIP} provides a finite-sample upper bound on PIP approximation errors in terms of the gap between the VR-IWAE objective and the log marginal likelihood. For fixed $\alpha$ and variational distribution, \eqref{eq:VR-IWAE-ordering} implies that this bound becomes no larger as $N$ increases. We subsequently use this result in Proposition~\ref{prop:collapsed-VR-IWAE} to establish that analytical marginalisation yields a tighter PIP error bound than the corresponding uncollapsed formulation.

We next turn to a stronger theoretical guarantee, investigating the approximation error of the globally optimised posterior inclusion probabilities and establishing a worst-case asymptotic rate of convergence under the following conditions.

\begin{assumption}
\label{ass:PIP-convergence}
The following conditions hold:
\begin{enumerate}
\item[(i)] The posterior satisfies
$p(\vgamma|\mathbf y)>0$ for every $\vgamma\in\{0,1\}^p$.

\item[(ii)] The variational probabilities are restricted to
$\boldsymbol{\rho}\in\mathcal R$, where
\[
\mathcal R
=
\left\{
\boldsymbol{\rho}\in\mathbb R^p:
\varepsilon\leq\rho_j\leq1-\varepsilon,
\quad j=1,\ldots,p
\right\},
\]
with
\[
0<\varepsilon<
\min_{1\leq j\leq p}
\left\{
\Pr(\gamma_j=1|\mathbf y),
1-\Pr(\gamma_j=1|\mathbf y)
\right\}.
\]

\item[(iii)] For each $N\geq1$ and $0<\alpha<1$,
the VR-IWAE objective in \eqref{eq:VR-IWAE-bound}
has a global maximiser over $[0,1]^p$ that lies
in $\mathcal R$, defined by
\begin{equation}
\widehat{\boldsymbol{\rho}}_{N,\alpha}
=
\underset{\boldsymbol{\rho}\in\mathcal R}{\arg\max}
\;
\ell_N^{(\alpha)}
\left(\operatorname{logit}(\boldsymbol{\rho})\right),
\label{eq:VR-IWAE-global-maximiser}
\end{equation}
where the logit transformation is applied componentwise.
\end{enumerate}

\end{assumption}

\begin{theorem}
\label{thm:PIP-convergence}
Under Assumption~\ref{ass:PIP-convergence}, for any fixed
$\alpha_0\in(0,1)$, there exists a constant $C>0$,
independent of $N$ and $\alpha$, such that
\begin{equation}
\max_{1\leq j\leq p}
\left|
\widehat\rho_{j,N,\alpha}
-
\Pr(\gamma_j=1|\mathbf y)
\right|
\leq
C\sqrt{\alpha+\frac{1}{N\alpha}},
\label{eq:optimal-PIP-bound}
\end{equation}
for all $N\geq1$ and $0<\alpha\leq\alpha_0$,
where $\widehat\rho_{j,N,\alpha}$ denotes the $j$th
coordinate of $\widehat{\boldsymbol{\rho}}_{N,\alpha}$ in \eqref{eq:VR-IWAE-global-maximiser}.
\end{theorem}

The following corollary follows immediately from
Theorem~\ref{thm:PIP-convergence}. Its proof is straightforward and omitted.
\begin{corollary}
\label{cor:PIP-rate}
Under Assumption~\ref{ass:PIP-convergence}, suppose that
$\alpha_N\asymp N^{-1/2}$. Then, as $N\to\infty$,
\begin{equation}
\max_{1\leq j\leq p}
\left|
\widehat\rho_{j,N,\alpha_N}
-
\Pr(\gamma_j=1|\mathbf y)
\right|
=
O(N^{-1/4}).
\label{eq:optimal-PIP-rate}
\end{equation}
\end{corollary}
The rate in Corollary~\ref{cor:PIP-rate} provides a worst-case upper bound on the PIP approximation error and does not preclude faster convergence in practice. Its main significance is therefore not the rate itself, but the guarantee that the optimised variational inclusion probabilities converge to their exact posterior counterparts without requiring independence among the inclusion indicators. As discussed in Section~\ref{sec:Introduction}, \citet{carbonetto2012scalable} attribute inaccurate PIPs under correlated predictors to their factorised variational approximation. Our result shows that this limitation need not persist when replacing the standard ELBO with an importance-weighted objective, providing a theoretical justification for retaining a scalable factorised variational family. Importantly, this convergence guarantee does not apply to the standard ELBO used in \citet{carbonetto2012scalable}, which corresponds to $N=1$ and therefore does not admit the same increasing-$N$ argument. Our numerical experiments in Section~\ref{sec:linear-regression-experiments} provide further empirical support for improved PIP estimation.

Finally, we show that integrating out the remaining parameters tightens the PIP error bound in Lemma~\ref{lem:VR-IWAE-PIP} for any finite $N$ relative to an uncollapsed formulation, providing a theoretical motivation for our collapsed approach. For the Gaussian regression model in Section~\ref{sec:collapsed-posterior}, let $\boldsymbol\psi=(\beta_0,\boldsymbol\beta_{\mathcal A},\sigma^2,\pi)$ collect the remaining parameters. Consider a joint variational distribution
\begin{equation}
q_{\vlambda}^{\mathrm J}(\vgamma,\boldsymbol\psi)
=
q_{\vlambda}(\vgamma)r(\boldsymbol\psi|\vgamma),
\label{eq:joint_variational}
\end{equation}
where $q_{\vlambda}$ is defined in \eqref{eq:variational-distribution} and $r(\boldsymbol\psi|\vgamma)$ is a conditional variational density. Let $\ell_{N,\mathrm J}^{(\alpha)}(\vlambda,r)$ denote the VR-IWAE objective defined analogously to \eqref{eq:VR-IWAE-bound}, but using the uncollapsed joint formulation $p(\mathbf y,\vgamma,\boldsymbol\psi)=p(\mathbf y|\vgamma,\boldsymbol\psi)p(\vgamma,\boldsymbol\psi)$ and the joint variational distribution $q_{\vlambda}^{\mathrm J}(\vgamma,\boldsymbol\psi)$.

\begin{proposition}
\label{prop:collapsed-VR-IWAE}
Suppose that the conditional variational density $r(\boldsymbol\psi|\vgamma)$ in \eqref{eq:joint_variational} has support containing that of $p(\boldsymbol\psi|\vgamma,\mathbf y)$. Then, for every $N\geq1$ and $0<\alpha<1$,
\begin{equation}
\ell_{N,\mathrm J}^{(\alpha)}(\vlambda,r)
\leq
\ell_N^{(\alpha)}(\vlambda)
\leq
\log p(\mathbf y).
\label{eq:collapsed-bound-inequality}
\end{equation}
Furthermore, for every predictor $j$,
\begin{equation}
\left|\Pr(\gamma_j=1|\mathbf y)-\rho_j\right|
\leq
\sqrt{\frac{\log p(\mathbf y)-\ell_N^{(\alpha)}(\vlambda)}{2\alpha}}
\leq
\sqrt{\frac{\log p(\mathbf y)-\ell_{N,\mathrm J}^{(\alpha)}(\vlambda,r)}{2\alpha}}.
\label{eq:collapsed-PIP-bound}
\end{equation}
\end{proposition}
Proposition~\ref{prop:collapsed-VR-IWAE} establishes that, for any finite $N$ and any admissible joint variational distribution with marginal $q_{\vlambda}(\vgamma)$, analytical marginalisation improves the VR-IWAE objective and provides an upper bound on PIP approximation errors that is at least as tight as that obtained from the uncollapsed formulation. This complements existing results for standard variational inference based on the KL divergence \citep{tseng2025variational}.

\subsection{Relationship to importance-sampling and post-processing}
\label{sec:postprocessing}

An alternative perspective on importance-weighted variational inference is to view $q_{\vlambda}$ as an importance-sampling proposal rather than solely as a variational approximation. 
For the ordinary IWAE objective ($\alpha=0$), \citet[Theorem~3]{Domke2018} imply that
\begin{equation}
\log p(\mathbf y)-\ell_N^{(0)}(\vlambda)
=
\frac{
\mathbb V_{q_{\vlambda}}
\left[
w_{\vlambda}(\vgamma;\mathbf y)
\right]
}{
2N p(\mathbf y)^2
}
+
o(N^{-1}),
\label{eq:IWAE-weight-variance}
\end{equation}
as $N\to\infty$.
Thus, asymptotically, maximising the IWAE objective corresponds to minimising the variance of the importance weights. A good importance distribution has low weight variance, which requires adequately covering regions where the posterior places substantial probability mass. This provides an intuitive explanation for the mass-covering behaviour of importance-weighted variational inference, in contrast to the mode-seeking behaviour of standard ELBO optimisation.

This importance-sampling interpretation motivates using the fitted variational distribution as a proposal for importance-sampling post-processing to improve posterior inference \citep{yao2018evaluating}. Marginalisation provides an additional advantage in this setting. To establish this, consider the joint variational distribution $q_{\vlambda}^{\mathrm J}$ in \eqref{eq:joint_variational} and define the corresponding joint importance weight
\begin{equation}
w_{\vlambda}^{\mathrm J}(\vgamma,\boldsymbol\psi;\mathbf y)
=
\frac{
p(\mathbf y,\vgamma,\boldsymbol\psi)
}{
q_{\vlambda}^{\mathrm J}(\vgamma,\boldsymbol\psi)
}.
\label{eq:joint-importance-weight}
\end{equation}
The following proposition shows that integrating out the remaining parameters reduces importance-weight variance and increases the asymptotic effective sample size.

\begin{proposition}
\label{prop:collapsed-IS}
Under the conditions of Proposition~\ref{prop:collapsed-VR-IWAE}, suppose additionally that the joint importance weights in \eqref{eq:joint-importance-weight} have finite variance. Then
\begin{equation}
\mathbb V_{q_{\vlambda}}
\left[
w_{\vlambda}(\vgamma;\mathbf y)
\right]
\leq
\mathbb V_{q_{\vlambda}^{\mathrm J}}
\left[
w_{\vlambda}^{\mathrm J}(\vgamma,\boldsymbol\psi;\mathbf y)
\right].
\label{eq:collapsed-variance-inequality}
\end{equation}
Furthermore, let $\operatorname{ESS}_M^{\mathrm C}$ and $\operatorname{ESS}_M^{\mathrm J}$ denote the effective sample sizes computed from $M$ independent collapsed and joint importance weights, respectively, where
\[
\operatorname{ESS}_M
=
\frac{
\left(\sum_{m=1}^M w_m\right)^2
}{
\sum_{m=1}^M w_m^2
}
\]
for unnormalised importance weights $w_1,\ldots,w_M$. Then
\begin{equation}
\lim_{M\to\infty}
\frac{\operatorname{ESS}_M^{\mathrm C}}{M}
\geq
\lim_{M\to\infty}
\frac{\operatorname{ESS}_M^{\mathrm J}}{M}.
\label{eq:collapsed-ESS-inequality}
\end{equation}
\end{proposition}

After the variational optimisation outlined in
Section~\ref{sec:stochastic-optimisation}, we use the
fitted variational distribution $q_{\widehat{\vlambda}}$
as an initial proposal for importance-sampling
post-processing. We refer to posterior estimates obtained
directly from the fitted variational distribution,
without post-processing, as \textit{original variational inference (VI)}. We distinguish
between non-adaptive and adaptive post-processing,
depending on whether the proposal distribution is
updated. 

Non-adaptive post-processing seeks to improve posterior
estimation by correcting discrepancies between the fitted
variational distribution and the posterior through importance
reweighting, without modifying the proposal or reoptimising
the variational objective. For $M$ independent draws
$\vgamma^{(1)},\ldots,\vgamma^{(M)}$ from
$q_{\widehat{\vlambda}}$, define the normalised importance
weights
\[
\widetilde w_m
=
\frac{
w_{\widehat{\vlambda}}(\vgamma^{(m)};\mathbf y)
}{
\sum_{k=1}^M
w_{\widehat{\vlambda}}(\vgamma^{(k)};\mathbf y)
}.
\]
The corresponding importance-sampling estimates of the
posterior inclusion probabilities are
\begin{equation}
\widehat{\rho}_j^{\,\mathrm{IS}}
=
\sum_{m=1}^M
\widetilde w_m\gamma_j^{(m)},
\qquad j=1,\ldots,p.
\label{eq:IS-PIP}
\end{equation}
To mitigate the influence of extreme importance weights,
we apply Pareto-smoothed importance sampling (PSIS)
\citep{vehtari2024pareto}, replacing the normalised
weights in \eqref{eq:IS-PIP} with their PSIS-smoothed
counterparts. We refer to this procedure as
\textit{ordinary PSIS}.

Adaptive post-processing seeks to improve importance-sampling reliability by updating the proposal distribution. This is motivated by the fact that accurate marginal PIPs do not necessarily guarantee reliable importance sampling, since the product-Bernoulli distribution cannot capture posterior dependence among inclusion indicators. We consider two approaches, both
starting from the same fitted $q_{\widehat{\vlambda}}$.
The first is a discrete adaptation of the
importance-weighted moment matching approach in
\citet{paananen2021implicitly}. In our setting, this
involves updating the variational Bernoulli probabilities
towards the PSIS-weighted inclusion probabilities,
obtained by replacing the normalised importance weights
in \eqref{eq:IS-PIP} with their Pareto-smoothed counterparts.
We refer to this procedure as \textit{importance-weighted
moment matching (IWMM)}. The second approach, which we
term \textit{iterative damped PSIS}, uses more conservative
updates towards the corresponding PSIS-weighted inclusion
probabilities, thereby limiting the magnitude of changes
to the proposal distribution. Both approaches draw new
model configurations from the adapted proposal and reweight
them to estimate posterior quantities, without reoptimising
the variational objective. Adaptive post-processing is optional, and the reliability of the resulting importance-sampling estimates is assessed using the Pareto $\widehat{k}$ diagnostic described in Section~\ref{sec:simulation-design}.

Implementation details of the post-processing methods are provided in
Section~\ref{sec:supp-postprocessing}.

\subsection{Gradient estimation and stochastic optimisation}
\label{sec:stochastic-optimisation}

We optimise the VR-IWAE objective in \eqref{eq:VR-IWAE-bound} using stochastic gradient ascent \citep{robbins1951stochastic}. Standard reparameterisation-based gradient estimators \citep{Kingma2013} can offer substantial variance reduction \citep{xu2019variance,domke2019provable}, but are not directly applicable in our setting because the inclusion indicators are discrete. We therefore adopt the score-function gradient estimators of \citet{daudel2026importance}, which extend the variational
inference for Monte Carlo objectives (VIMCO) estimator originally proposed by \citet{mnih2016variational} to the VR-IWAE framework.

For the product-Bernoulli variational distribution in \eqref{eq:variational-distribution}, the score function has the simple form
\begin{equation}
s_j(\vgamma)
=
\frac{\partial}{\partial\lambda_j}
\log q_{\vlambda}(\vgamma)
=
\gamma_j-\rho_j,
\qquad j=1,\ldots,p.
\label{eq:bernoulli-score}
\end{equation}
To estimate the gradient, we generate $S$ independent groups of $N$ samples from $q_{\vlambda}$, denoted by $\vgamma_i^{(s)}$, $i=1,\ldots,N$ and $s=1,\ldots,S$. Define the tempered importance weights and their normalised counterparts as
\[
w_{si}^{(\alpha)}
=
w_{\vlambda}(\vgamma_i^{(s)};\mathbf y)^{1-\alpha},
\qquad
\widetilde w_{si}^{(\alpha)}
=
\frac{w_{si}^{(\alpha)}}
{\sum_{\ell=1}^N w_{s\ell}^{(\alpha)}}.
\]
The corresponding Monte Carlo estimate of the VR-IWAE objective in \eqref{eq:VR-IWAE-bound} for group $s$ is
\[
\widehat{\ell}_{N,s}^{(\alpha)}(\vlambda)
=
\frac{1}{1-\alpha}
\log\left(
\frac{1}{N}\sum_{i=1}^N w_{si}^{(\alpha)}
\right).
\]
A direct application of the score-function identity yields the naive gradient estimator
\[
\widehat g_j^{(\alpha,\mathrm{naive})}
=
\frac{1}{S}
\sum_{s=1}^S\sum_{i=1}^N
s_j(\vgamma_i^{(s)})
\left(
\widehat{\ell}_{N,s}^{(\alpha)}(\vlambda)
-
\widetilde w_{si}^{(\alpha)}
\right).
\]
\citet{daudel2026importance} show that, for fixed $0<\alpha<1$ and under suitable regularity conditions, the variance of this estimator is of order $O(N/S)$, while its signal-to-noise ratio (SNR) is of order $O(\sqrt{S/N})$. Here, the SNR of a gradient estimator $G$ is defined as $|\E[G]|/\sqrt{\mathbb V(G)}$. This creates a tension: increasing $N$ tightens the VR-IWAE objective but also increases the variance and decreases the SNR of the naive gradient estimator, making optimisation
increasingly challenging.

To address this issue, we adopt the VIMCO approach of \citet{mnih2016variational}, as extended by \citet{daudel2026importance} to the VR-IWAE framework. For $N\geq2$, the general VIMCO gradient estimator takes the form
\begin{equation}
\widehat g_j^{(\alpha)}
=
\frac{1}{S}
\sum_{s=1}^S\sum_{i=1}^N
s_j(\vgamma_i^{(s)})
\left(
\widehat{\ell}_{N,s}^{(\alpha)}(\vlambda)
-
\widetilde w_{si}^{(\alpha)}
-
b_{-i,s,j}^{(\alpha)}
\right),
\label{eq:VIMCO-gradient}
\end{equation}
where $b_{-i,s,j}^{(\alpha)}$ is a control variate constructed independently of $\vgamma_i^{(s)}$, ensuring that the gradient estimator remains unbiased. For fixed $0<\alpha<1$, the VIMCO estimators considered below have asymptotic variance of order $O(1/(SN))$ and SNR of order $O(\sqrt{SN})$ \citep{daudel2026importance}. Thus, unlike the naive estimator, the VIMCO estimators benefit from increasing the number of importance samples. We consider two control-variate choices proposed by \citet{daudel2026importance}. The first, VIMCO-$\star$, uses an asymptotically variance-minimising control variate, which we estimate using samples generated at the previous optimisation iteration. The second, VIMCO-GM, uses a geometric-mean (GM) control variate that is computationally inexpensive. Both estimators optimise the same VR-IWAE objective in \eqref{eq:VR-IWAE-bound} and differ only in the construction of their control variates. Their explicit expressions and implementation details are provided in Section~\ref{sec:appendix:Estimating the Control Variates}.

A further consideration is the choice of the annealing parameter $\alpha$. Larger values of $\alpha$ stabilise the tempered importance weights and can reduce the variability of the stochastic gradients, whereas smaller values yield a tighter VR-IWAE objective. We therefore consider an annealing strategy that initialises $\alpha$ close to one and gradually decreases it as the variational approximation improves. We adopt the ESS-based annealing strategy in \citet{daudel2026importance}, extending their criterion to $S$ independent groups by summing the group-specific effective sample sizes:
\begin{equation}
\operatorname{ESS}_{1-\alpha}
=
\sum_{s=1}^S
\frac{
\left(\sum_{i=1}^N w_{si}^{(\alpha)}\right)^2
}{
\sum_{i=1}^N \left(w_{si}^{(\alpha)}\right)^2
}.
\label{eq:annealing-ESS}
\end{equation}
The parameter $\alpha$ is progressively decreased whenever $\operatorname{ESS}_{1-\alpha}$ exceeds a prescribed threshold, allowing the VR-IWAE objective to be tightened as the importance weights become more balanced. The choices of $N$ and $S$ also involve a trade-off: increasing $N$ tightens the objective, whereas increasing $S$ reduces gradient variability through independent averaging. For a fixed computational budget $SN$, the asymptotic results in \citet{daudel2026importance} favour a larger $N$ and smaller $S$. 

Table~\ref{tab:methods-overview} summarises the methods considered in this paper, distinguishing between original variational optimisation and the optional subsequent importance-sampling post-processing. The table also serves as a reference for the corresponding method abbreviations.

\begin{table}[h]
\centering
\small
\caption{Overview of the original variational inference and subsequent importance-sampling post-processing methods. Abbreviations: ELBO, evidence lower bound; VIMCO, variational inference for Monte Carlo objectives; GM, geometric mean; PSIS, Pareto-smoothed importance sampling; IWMM, importance-weighted moment matching.}
\label{tab:methods-overview}
\begin{tabular}{llp{0.45\linewidth}}
\toprule
Method & Component & Description \\
\midrule
ELBO & Optimisation & Standard variational baseline \\
VIMCO-$\star$ & Optimisation & Fixed $\alpha$, variance-minimising control variate \\
VIMCO-GM & Optimisation & Fixed $\alpha$, geometric-mean control variate \\
Annealed VIMCO-$\star$ & Optimisation & VIMCO-$\star$ with adaptive $\alpha$ \\
Annealed VIMCO-GM & Optimisation & VIMCO-GM with adaptive $\alpha$ \\
PSIS & Post-processing & Pareto smoothing of importance weights \\
Discrete IWMM & Post-processing & Adaptation of Bernoulli probabilities \\
Iterative Damped PSIS & Post-processing & Damped proposal adaptation \\
\bottomrule
\end{tabular}
\end{table}

\section{Experiments}\label{sec:linear-regression-experiments}

\subsection{Experimental setup}
\label{sec:experimental-setup}

All experiments are conducted on an Apple MacBook Pro equipped with an Apple M2 processor (8 CPU cores) and 8 gigabytes of RAM.

\begin{itemize}
\item \textbf{Experiment 1: Posterior inclusion probability accuracy.}
We assess posterior inclusion probability accuracy against collapsed Gibbs sampling references for $p=30$ and $p=1000$, with the latter providing a finite-budget approximation. We also examine the reliability of importance-sampling post-processing in both settings.

\item \textbf{Experiment 2: High-dimensional variable selection and uncertainty quantification.}
For $p=5000$, where a suitable Gibbs-based posterior inclusion probability reference is unavailable, we investigate variable-selection accuracy, uncertainty quantification and importance-sampling post-processing reliability.

\item \textbf{Experiment 3: Comparison with existing variational approaches.}
Using the $p=30$ setting from Experiment 1, we compare the posterior inclusion probability accuracy of the original variational approximations in this paper with that of the methods proposed in \citet{carbonetto2012scalable} and \citet{you2023approximate}. Each method is evaluated against a Gibbs sampling reference under its corresponding model formulation, without importance-sampling post-processing.
\end{itemize}

\subsection{Experimental settings and evaluation metrics}
\label{sec:simulation-design}
All variational methods other than those in \citet{carbonetto2012scalable} and \citet{you2023approximate} follow the optimisation, tuning and post-processing protocol detailed in Section~\ref{sec:supp-implementation}.

For each simulation setting, we generate $R=50$ independent datasets. Experiment 3 reuses the $p=30$ datasets from Experiment 1. The design is adapted from \citet{carbonetto2012scalable}, replacing their genotype-motivated covariates with a correlated Gaussian AR(1) design \citep{leng2014bayesian},
\[
\operatorname{Corr}(X_j,X_k)=0.8^{|j-k|}.
\]
This local dependence is relevant because factorised variational approximations may distort posterior inclusion probabilities under correlated predictors. Responses are generated according to
\[
\mathbf y=\beta_0\mathbf 1_n+\mathbf X\boldsymbol\beta+\boldsymbol\varepsilon,
\qquad
\boldsymbol\varepsilon\sim N(\mathbf 0,9\mathbf I_n),
\qquad \beta_0=3.
\]
The three settings are summarised in Table~\ref{tab:simulation-design}. The inclusion prior is calibrated so that its expected model size approximately matches the true sparsity. Within each setting, the non-zero coefficients are drawn once from $N(0,1)$ and held fixed across replications.

\begin{table}[H]
\centering
\caption{Simulation settings for the three dimensionality scenarios.}
\label{tab:simulation-design}
\begin{tabular}{rrrr}
\toprule
$n$ & $p$ & Active predictors & Prior on $\pi$ \\
\midrule
1000 & 30   & 5  & $\operatorname{Beta}(0.2,1)$ \\
500  & 1000 & 20 & $\operatorname{Beta}(0.02,1)$ \\
500  & 5000 & 20 & $\operatorname{Beta}(0.004,1)$ \\
\bottomrule
\end{tabular}
\end{table}

For $p=30$ and $p=1000$, we assess posterior inclusion probability accuracy using PIP root mean squared error (RMSE) and the mean marginal Kullback--Leibler (KL) discrepancy, defined by
\[
\begin{aligned}
\operatorname{RMSE}_{\mathrm{PIP}}
&= \left\{\frac{1}{p}\sum_{j=1}^{p}
\left(\widehat\rho_j-\rho_j^\star\right)^2\right\}^{1/2},
\\
D_{\mathrm{PIP}}
&= \frac{1}{p}\sum_{j=1}^{p}
\operatorname{KL}\!\left(
\operatorname{Bernoulli}(\rho_j^\star)
\,\middle\|\,
\operatorname{Bernoulli}(\widehat\rho_j)
\right),
\end{aligned}
\]
where $\rho_j^\star$ and $\widehat\rho_j$ are the reference and estimated posterior inclusion probabilities, respectively. The reference is obtained by collapsed Gibbs sampling, which provides an accurate reference for $p=30$ and a finite-budget approximation based on 100,000 iterations for $p=1000$. At $p=5000$, where a suitable Gibbs PIP reference is unavailable, we instead report the $F_1$ score (the harmonic mean of precision and recall)
 for recovery of the true active predictors, selecting predictors with estimated PIPs exceeding 0.5, and the interval score for central 95\% credible intervals of the latent signal $f_\star=\beta_0+\mathbf x_\star^\top\boldsymbol\beta$. The latter is evaluated at 100 new predictor vectors, for which the true signal is known. We construct these intervals by drawing inclusion vectors from the variational approximations and averaging the corresponding conditional posterior distributions, with the regression coefficients and residual variance integrated out analytically.

We assess importance-sampling reliability using the Pareto shape diagnostic $\widehat{k}$ for ordinary PSIS, IWMM and iterative damped PSIS \citep{vehtari2024pareto}. The adaptive post-processing procedures seek to reduce $\widehat{k}$ by updating the proposal distribution. We report the proportion of fits with a defined $\widehat{k}\leq0.7$ for each procedure and variational method, distinguishing fits for which $\widehat{k}$ is undefined. 
An undefined $\widehat{k}$ may arise when repeated draws of the
same inclusion vector produce identical importance ratios,
preventing a reliable fit to their upper tail, and does not
necessarily indicate a poor posterior approximation. Moreover, $\widehat{k}$ diagnoses importance-sampling reliability rather than posterior approximation accuracy, so a smaller value does not necessarily imply more accurate posterior inclusion probabilities \citep{yao2018evaluating,vehtari2024pareto}.

For Experiments 1 and 2, we present the results through three comparisons of inferential accuracy and importance-sampling reliability. First, we report within-dataset paired differences between the original VIMCO approximations and the original ELBO to evaluate the effects of importance-weighted optimisation on inferential performance. Second, we compare ordinary PSIS with the corresponding original
variational approximation to evaluate the effects of
non-adaptive post-processing on inferential performance. Third, we examine whether proposal adaptation improves importance-sampling reliability relative to ordinary PSIS. For the first two comparisons, paired differences in the
performance metrics are defined so that positive values indicate an improvement over the baseline. We report mean paired differences with 95\% Monte Carlo confidence intervals based on the 50 independent datasets. Table~\ref{tab:simulation-runtime} reports the mean runtime of the selected variational optimisation, excluding tuning and post-processing. Table~\ref{tab:postprocess-ess} reports the median effective sample size (ESS) of the post-processing importance weights. 
Table~\ref{tab:supp-alpha-iter-sim} reports the mean selected or terminal values of $\alpha$ and the mean number of optimisation iterations.

\begin{table}[h]
\centering
\caption{Mean optimisation runtime (seconds) for the selected variational configuration in the three simulation settings. The timings exclude tuning and importance-sampling post-processing.}
\label{tab:simulation-runtime}
\begin{tabular}{lrrr}
\toprule
Method & $p=30$ & $p=1000$ & $p=5000$ \\
\midrule
ELBO & 0.7750 & 35.2968 & 174.5873 \\
VIMCO-$\star$ & 0.8969 & 15.3979 & 86.5798 \\
Annealed VIMCO-$\star$ & 1.1217 & 16.0699 & 80.8916 \\
VIMCO-GM & 0.8772 & 9.6514 & 49.9620 \\
Annealed VIMCO-GM & 0.8665 & 10.7708 & 41.2546 \\
\bottomrule
\end{tabular}
\end{table}

In Experiment 3, we compare the variational approximations
considered in this paper, without importance-sampling
post-processing, with the methods of \citet{carbonetto2012scalable} and \citet{you2023approximate}, using the same $p=30$ datasets as Experiment 1. Unlike our collapsed formulation, \citet{carbonetto2012scalable} approximate the regression coefficients and inclusion indicators jointly, conditional on model hyperparameters. Their method is implemented in the \texttt{varbvs} package in \texttt{R}, which estimates the residual and slab variances over a grid of prior inclusion odds and averages PIPs using fitted hyperparameter weights. In contrast, the approximate collapsed variational Bayes (ACVB) method of \citet{you2023approximate} integrates out the regression coefficients and residual variance before approximating the posterior over inclusion indicators. We run ACVB from multiple initialisations and retain the solution with the highest approximate objective. Because the model formulations differ, we assess each method against a Gibbs sampling reference under its own formulation. For \texttt{varbvs}, the Gibbs reference uses the same fitted hyperparameters and weights, isolating the error due to the mean-field approximation. For ACVB, the Gibbs reference targets its corresponding exact collapsed posterior. We restrict this comparison to $p=30$, where reliable posterior-specific Gibbs references are computationally feasible.

\subsection{Experiment 1: Posterior inclusion probability accuracy}
\label{sec:p30-1000results}

\begin{figure}[h]
\centering
\includegraphics[width=\linewidth]{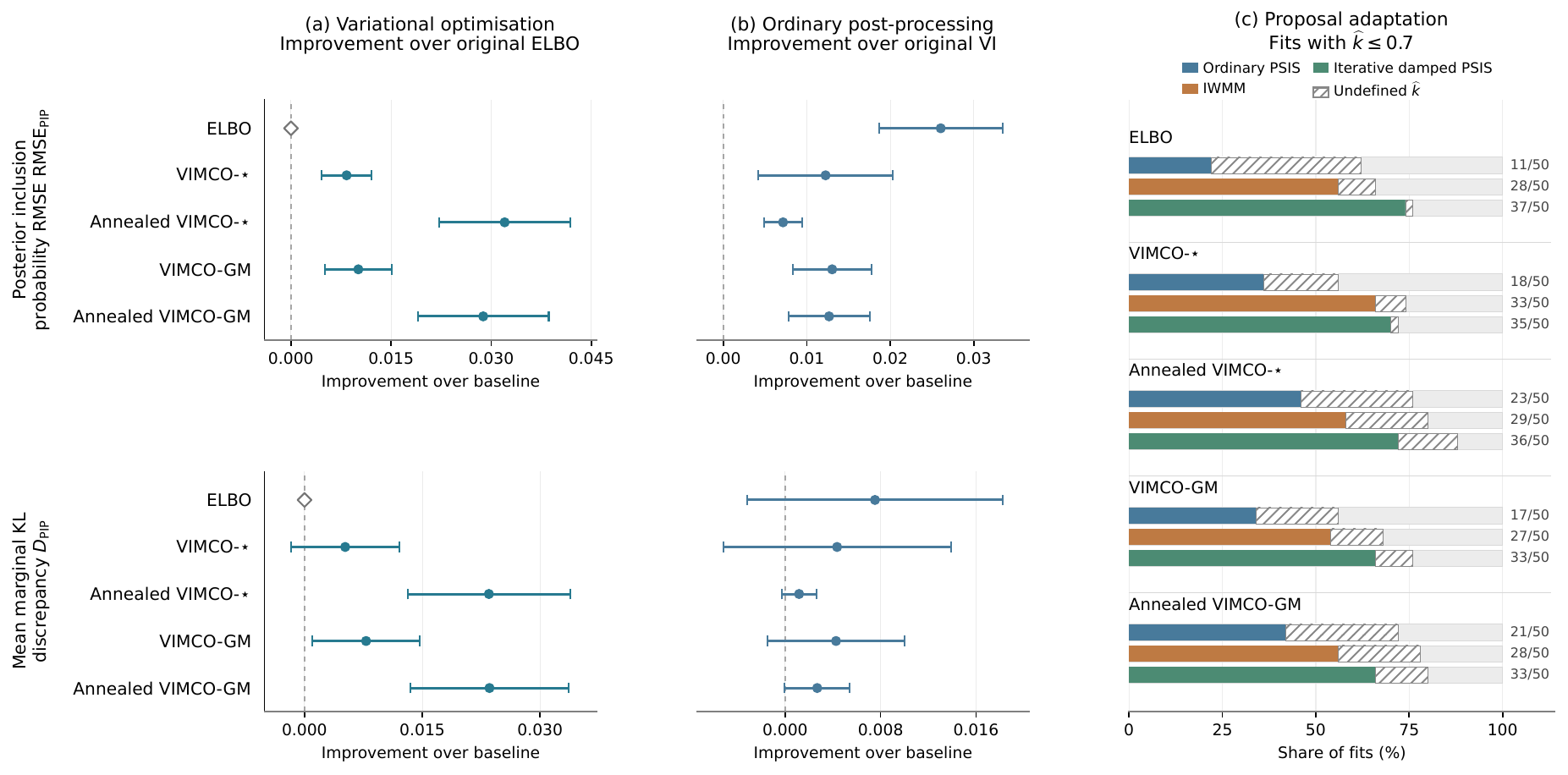}
\caption{Posterior inclusion probability (PIP) accuracy and importance-sampling reliability for $p=30$ across 50 independently simulated datasets. Panels (a) and (b) show mean within-dataset improvements in PIP RMSE and mean marginal KL discrepancy relative to the original ELBO and the corresponding original variational approximation, respectively; positive values indicate improved accuracy. Error bars are 95\% Monte Carlo confidence intervals for paired differences. Panel (c) reports the percentage of all 50 fits with defined Pareto $\widehat{k}\leq0.7$ for ordinary PSIS, IWMM and iterative damped PSIS. Undefined diagnostics are indicated separately, and the remaining percentage has defined $\widehat{k}>0.7$.}
\label{fig:vimco-abc-p30}
\end{figure}

\begin{figure}[h]
\centering
\includegraphics[width=\linewidth]{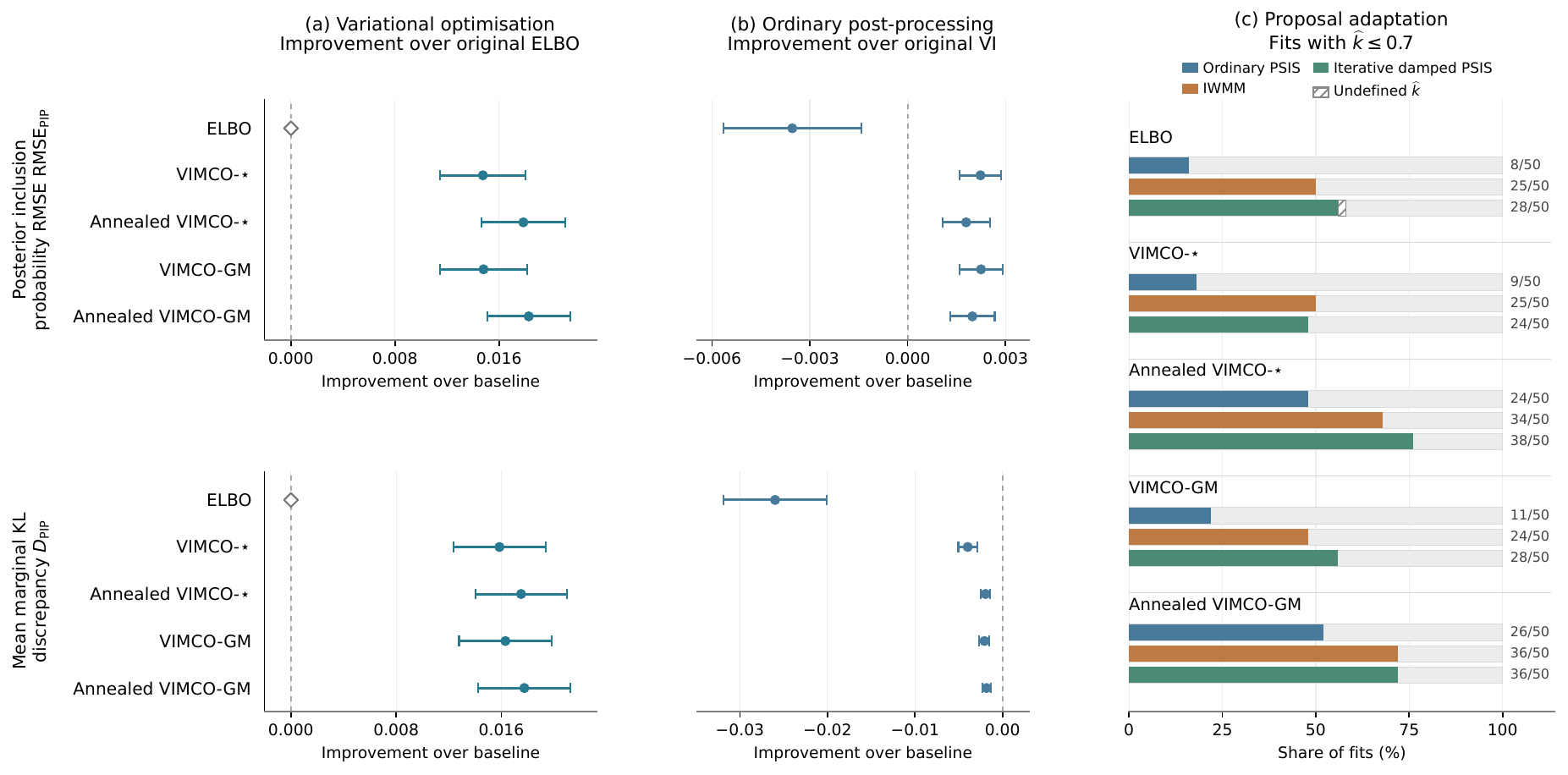}
\caption{Posterior inclusion probability (PIP) accuracy and importance-sampling reliability for $p=1000$.  The panel definitions and plotting details are as in Figure~\ref{fig:vimco-abc-p30}.}
\label{fig:vimco-abc-p1000}
\end{figure}

Figures~\ref{fig:vimco-abc-p30} and~\ref{fig:vimco-abc-p1000} present the results on posterior inclusion probability accuracy and importance-sampling reliability for $p=30$ and $p=1000$, respectively. Panels (a) show that all four VIMCO variants improve PIP RMSE relative to the original ELBO at both dimensions, with particularly consistent improvements across methods at $p=1000$ and generally similar improvements in mean marginal KL discrepancy. Annealing provides particularly pronounced gains at $p=30$, while the additional benefits are less pronounced at $p=1000$, where annealing terminates at a higher value of $\alpha$, consistent with more severe importance-weight degeneracy; see Table~\ref{tab:supp-alpha-iter-sim}. Panels (b) show that ordinary PSIS improves PIP RMSE for all methods at $p=30$. At $p=1000$, it improves PIP RMSE for the VIMCO methods but worsens both accuracy measures for ELBO, while mean marginal KL discrepancy increases for all methods. These accuracy results should be interpreted jointly with the importance-sampling reliability diagnostics in panels (c). In particular, the deterioration for ELBO is consistent with its poor importance-sampling reliability, with only 16\% of fits satisfying $\widehat{k}\leq 0.7$ under ordinary PSIS. This illustrates the limitations of importance reweighting when the variational proposal does not sufficiently capture the posterior distribution, particularly in high dimensions. Panels (c) moreover show that all four VIMCO proposals generally yield more favourable importance-sampling diagnostics than ELBO, with proposal adaptation further improving reliability in many cases. Undefined $\widehat{k}$ estimates occur more frequently at $p=30$, where repeated draws of the same discrete models can leave insufficient variation in the upper tail of the importance weights for Pareto fitting. 

\FloatBarrier

\begin{figure}[h]
\centering
\includegraphics[width=\linewidth]{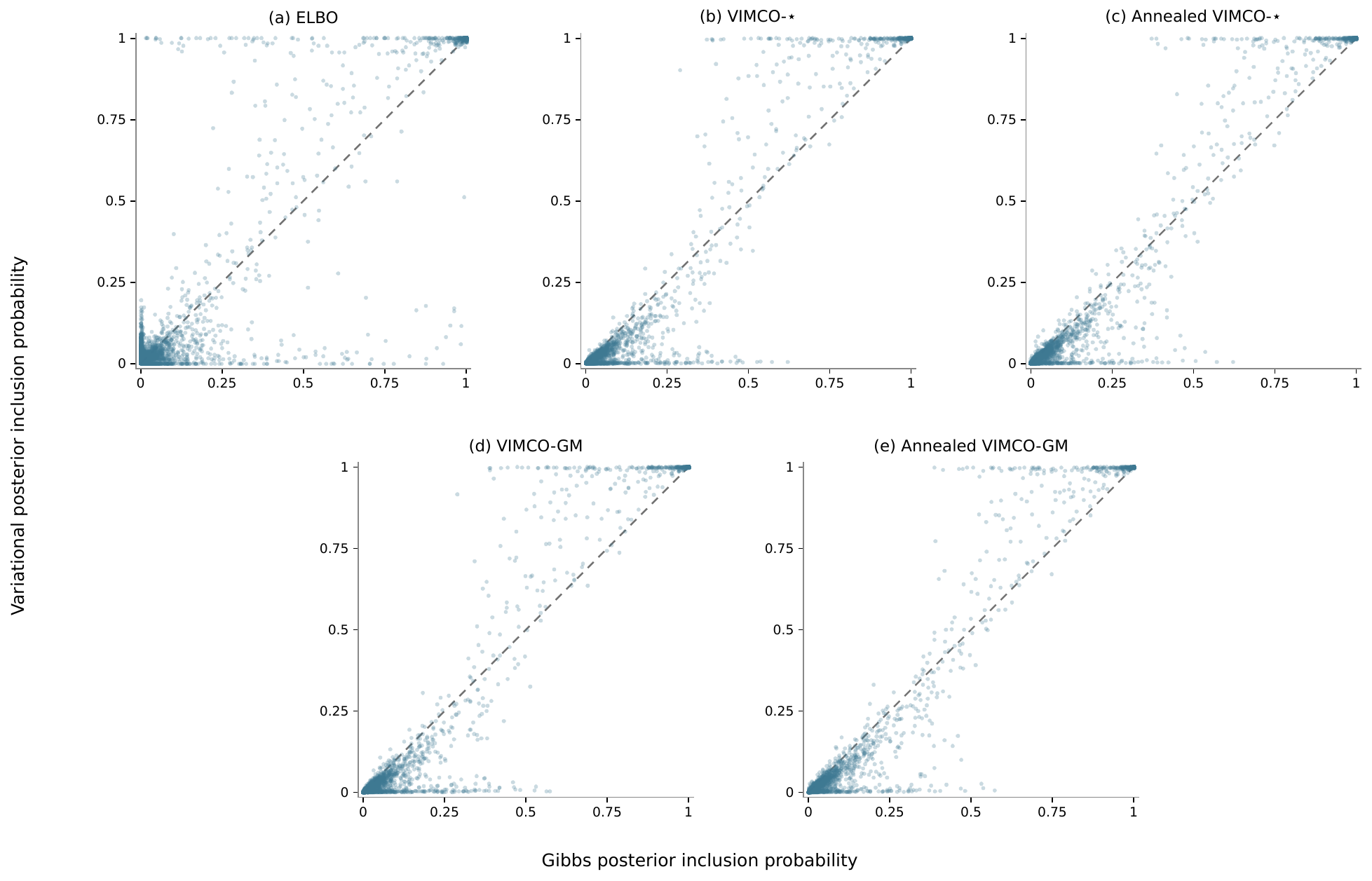}
\caption{Predictor-level posterior inclusion probabilities from the five original variational approximations versus collapsed Gibbs PIP references at $p=1000$. Each panel pools all 1,000 predictors from each of the 50 independently simulated datasets (50,000 points per method). The dashed diagonal denotes agreement with the reference. The Gibbs PIPs are finite-budget estimates based on 100,000 iterations. No importance-sampling post-processing is applied.}
\label{fig:vimco-pips-p1000-original}
\end{figure}

Figure~\ref{fig:vimco-pips-p1000-original} compares the original variational PIPs with the Gibbs reference at $p=1000$. The ELBO exhibits substantial deviations from the diagonal, including many PIPs close to zero or one when the Gibbs reference does not strongly favour either inclusion or exclusion. These discrepancies are markedly reduced by the VIMCO variants, particularly when combined with annealing, which yield closer agreement with the Gibbs reference. These findings are consistent with Corollary~\ref{cor:PIP-rate}, which establishes an upper bound on the PIP approximation error that decreases as the number of importance samples $N$ increases, with $\alpha$ approaching zero appropriately. In contrast, the standard ELBO has no analogous convergence guarantee. Importantly, these improvements are achieved directly through importance-weighted optimisation, without importance-sampling post-processing.

Overall, Experiment 1 demonstrates that importance-weighted optimisation improves PIP accuracy relative to the standard ELBO, while also yielding proposals with more favourable importance-sampling diagnostics. These importance-weighted improvements do not come at the expense of computational efficiency. 
All variational methods use the same per-iteration budget of 300 Monte
Carlo samples; see Section~\ref{sec:supp-implementation}. Table~\ref{tab:simulation-runtime} shows that optimisation times are comparable at $p=30$, whereas at $p=1000$ all VIMCO variants are faster than ELBO, with the VIMCO-GM variants requiring approximately 10--11 seconds compared with 35 seconds for ELBO. Additional results, including PIP comparisons before and after post-processing and the effective sample sizes of the importance weights, are provided in Section~\ref{sec:additional-experiment-results}. Table~\ref{tab:supp-alpha-iter-sim} additionally reports optimisation iteration counts across all simulation settings.

\FloatBarrier

\subsection{Experiment 2: High-dimensional variable selection and uncertainty quantification}
\label{sec:p5000results}

\begin{figure}[h]
\centering
\includegraphics[width=\linewidth]{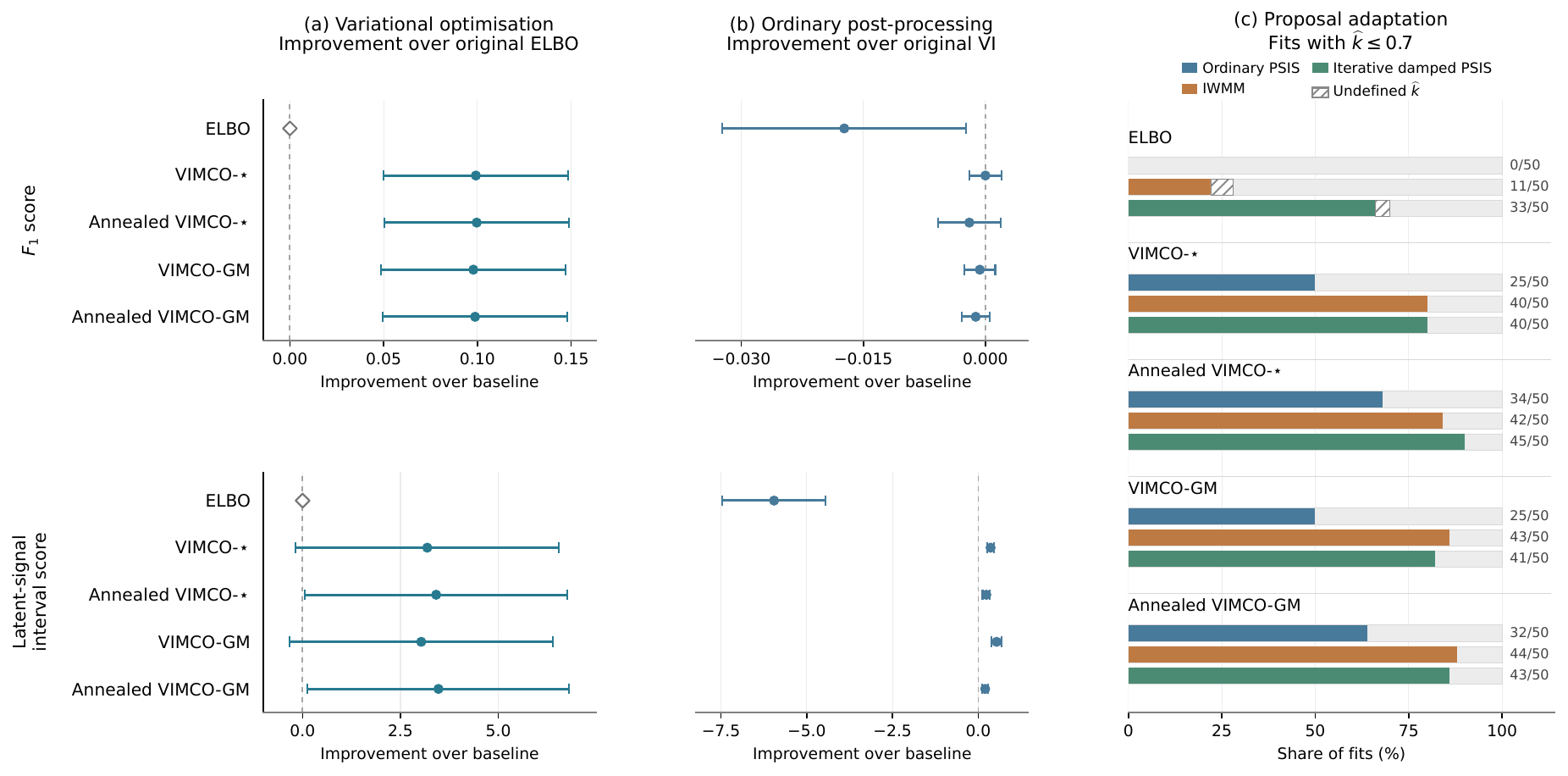}
\caption{Variable-selection accuracy, latent-signal uncertainty quantification and importance-sampling reliability at $p=5000$ across 50 independently simulated datasets. Panels (a) and (b) report mean paired improvements over the original ELBO and the corresponding original variational approximation in $F_1$ score and the 95\% latent-signal interval score; positive values denote improvements. Error bars are 95\% Monte Carlo confidence intervals. Panel (c) reports the percentage of all 50 fits with defined Pareto $\widehat{k}\leq0.7$ for ordinary PSIS, IWMM and iterative damped PSIS, with undefined diagnostics shown separately.}
\label{fig:vimco-abc-p5000}
\end{figure}

Figure~\ref{fig:vimco-abc-p5000} presents the results on variable-selection accuracy, latent-signal uncertainty quantification and importance-sampling reliability at $p=5000$. This is a particularly challenging high-dimensional setting, with 5,000 candidate predictors, only 500 observations and an enormous space of possible models. Despite this complexity, panel (a) shows that all four VIMCO variants improve the $F_1$ score relative to ELBO, with average improvements of approximately 0.10. Improvements in the 95\% latent-signal interval score are also positive on average, although the confidence intervals are wider. Panel (b) shows that ordinary PSIS provides little additional improvement for the VIMCO methods, whereas it worsens both accuracy measures for ELBO. These findings should be interpreted alongside panel (c), where none of the ELBO fits satisfies $\widehat{k}\leq0.7$ under ordinary PSIS, compared with 50--68\% of fits for the VIMCO proposals. Proposal adaptation substantially improves importance-sampling reliability, although the extent of the improvement varies across methods.

\FloatBarrier

Table~\ref{tab:simulation-runtime} shows that the computational advantages of VIMCO become more pronounced at $p=5000$. The ELBO requires approximately 175 seconds for optimisation, compared with 41--87 seconds for the VIMCO variants, with VIMCO-GM and annealed VIMCO-GM being the fastest; see also Table~\ref{tab:supp-alpha-iter-sim} for the corresponding optimisation iteration counts. The corresponding effective sample sizes of the importance weights are reported in Table~\ref{tab:postprocess-ess}.

\FloatBarrier

\subsection{Experiment 3: Comparison with existing variational approaches}
\label{sec:comparison_carbonetto_stephens_results}

\begin{figure}[htbp]
\centering
\includegraphics[width=\linewidth]{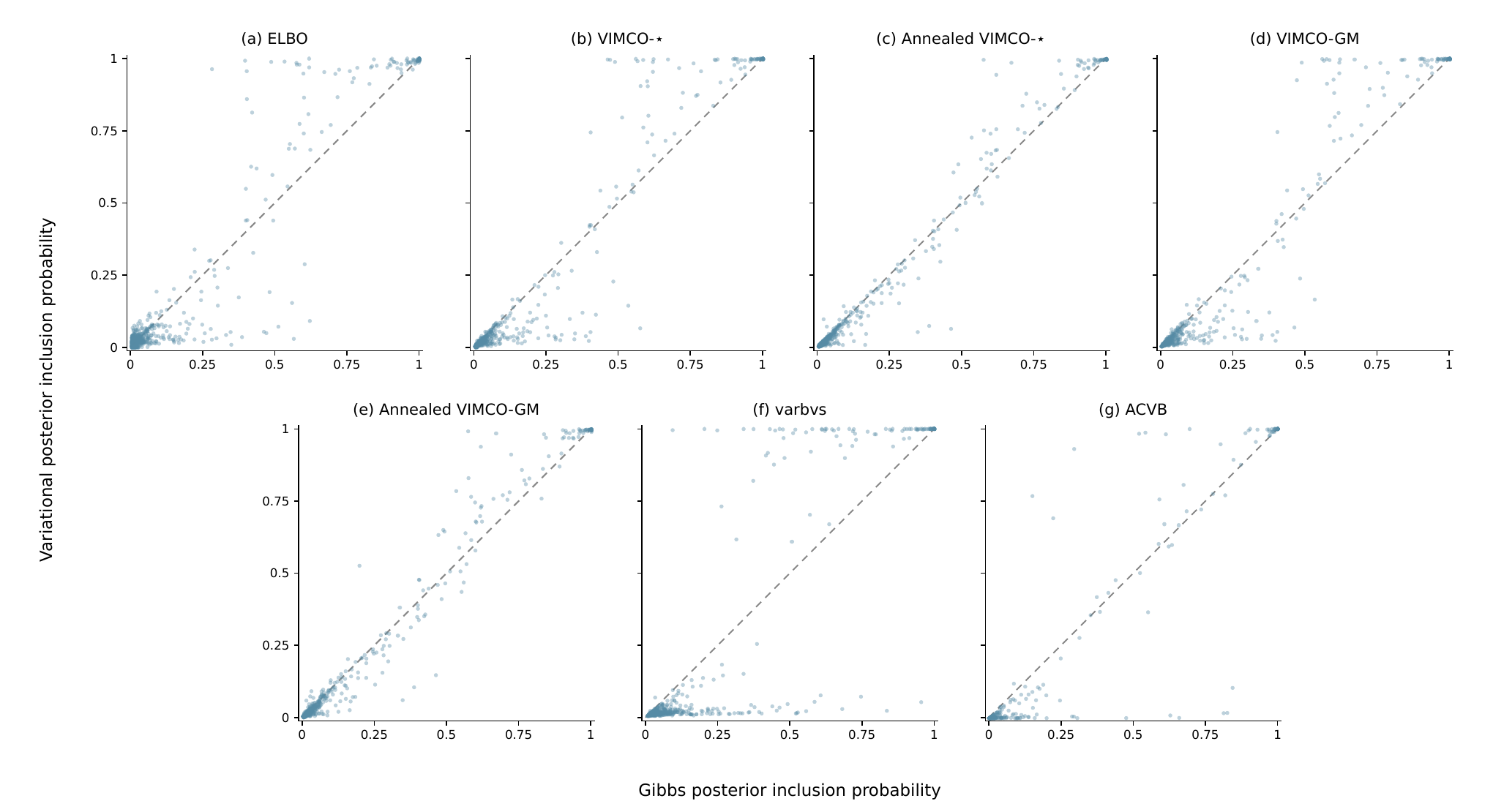}
\caption{Predictor-level posterior inclusion probabilities from the five original variational approximations, \texttt{varbvs}, and approximated collapsed variational Bayes (ACVB) versus their corresponding Gibbs references at $p=30$. Each panel pools 30 predictors across the same 50 simulated datasets (1,500 points per method). Our methods are evaluated against the collapsed Gibbs reference under our formulation, without importance-sampling post-processing; \texttt{varbvs} uses a Gibbs reference with its fitted hyperparameters and averaging weights; and ACVB uses a Gibbs reference targeting its corresponding collapsed posterior. The dashed diagonal denotes agreement with the respective reference.}
\label{fig:vimco-pips-p30-varbvs}
\end{figure}

Figure~\ref{fig:vimco-pips-p30-varbvs} compares predictor-level posterior inclusion probabilities from the original variational approximations in this paper with those from \texttt{varbvs} \citep{carbonetto2012scalable} and the approximated collapsed variational Bayes (ACVB, \citealp{you2023approximate}). Each method is evaluated against a Gibbs sampling reference under its own model formulation, and the comparison is based on the visual agreement between the variational posterior inclusion probabilities and the corresponding Gibbs references. The VIMCO methods exhibit noticeably closer agreement with their corresponding Gibbs references than both \texttt{varbvs} and ACVB, particularly when combined with annealing. The standard ELBO under our collapsed formulation also performs competitively. ACVB appears to perform marginally better than \texttt{varbvs}, which is consistent with the benefits of marginalisation for variational inference discussed in \citet{you2023approximate}.

\FloatBarrier

\section{Applications}\label{sec:Applications}

\subsection{Body fat data}
\label{sec:application-bodyfat}

The body fat dataset contains measurements on 252 men, with percentage body fat as the response and 13 candidate predictors, including age, weight, height and ten body-circumference measurements \citep{johnson1996, hoeting1999}. Following \citet{hoeting1999}, we remove one observation with an erroneous recorded height, leaving $n=251$. Motivated by previous Bayesian model-averaging analyses, we specify $\pi\sim\operatorname{Beta}(0.3,1)$, corresponding to a prior expected model size of approximately three. With only 13 predictors, all $2^{13}=8192$ models can be enumerated, providing exact posterior inclusion probabilities against which to evaluate our variational approximations.

\begin{figure}[h]
\centering
\includegraphics[width=\linewidth]{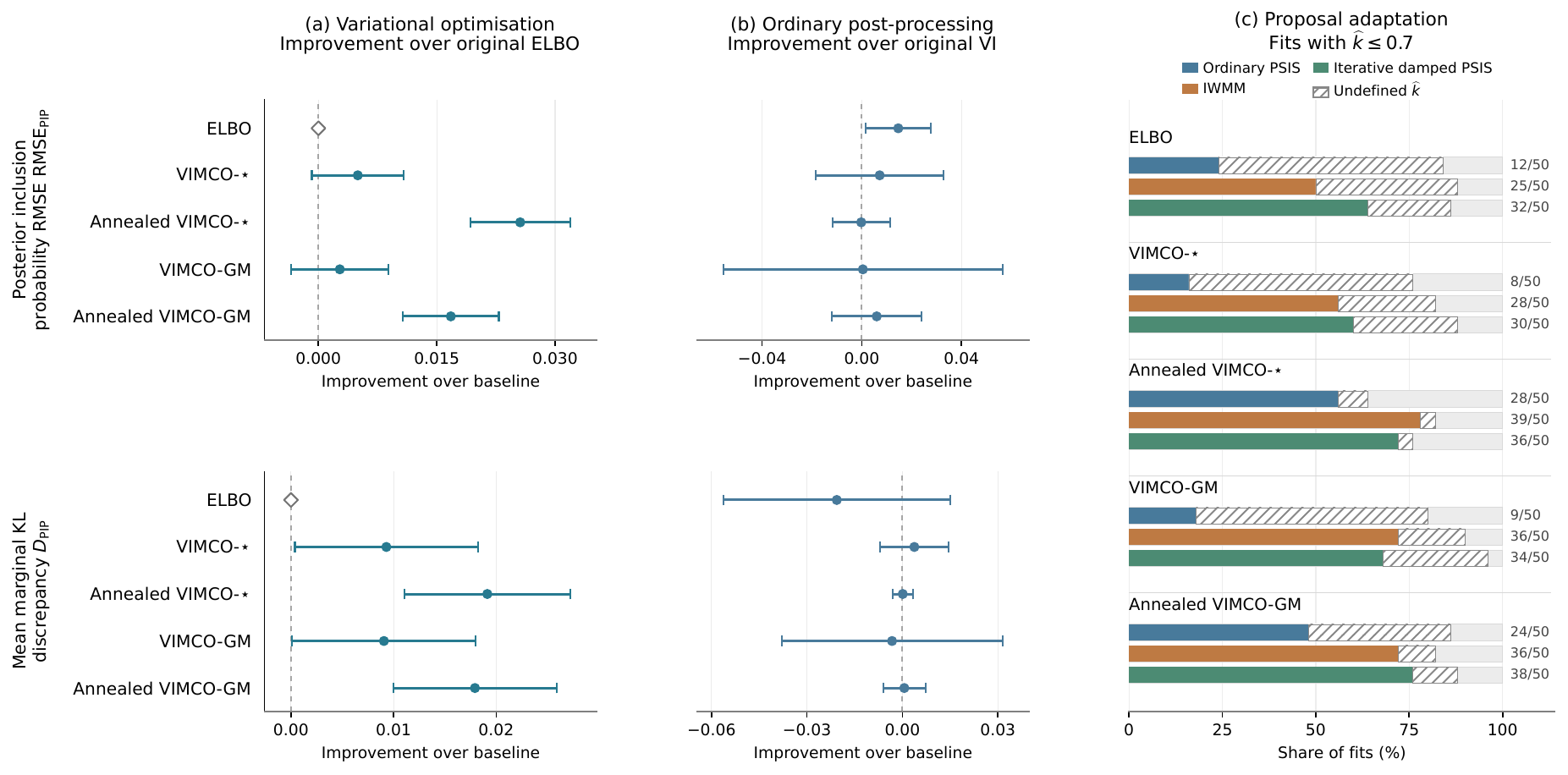}
\caption{Posterior inclusion probability (PIP) accuracy and importance-sampling reliability for the body fat application. PIP accuracy is evaluated against the exact posterior obtained by model enumeration. Panels (a) and (b) display mean within-optimisation run improvements in PIP RMSE and mean marginal KL discrepancy relative to the original ELBO and the corresponding original variational approximation, respectively; positive values indicate improved accuracy. Error bars represent one standard deviation across 50 optimisation runs on the same dataset and are not confidence intervals. Panel (c) reports the percentage of all 50 fits with defined Pareto $\widehat{k}\leq0.7$ for ordinary PSIS, IWMM and iterative damped PSIS. Undefined diagnostics are indicated separately, and the remaining percentage has defined $\widehat{k} > 0.7$}
\label{fig:vimco-abc-bodyfat}
\end{figure}

Figure~\ref{fig:vimco-abc-bodyfat} summarises posterior inclusion probability accuracy and importance-sampling reliability across the 50 optimisation fits. Panel (a) shows that all four VIMCO variants improve PIP RMSE and mean marginal KL discrepancy relative to ELBO, with particularly pronounced gains from annealing. Panel (b) shows that the effects of ordinary PSIS on PIP approximation accuracy are less consistent, with substantial variation across optimisation fits making it difficult to draw general conclusions. The importance-sampling reliability diagnostics in panel (c), however, favour the annealed VIMCO variants. Under ordinary PSIS, 56\% and 48\% of fits for annealed VIMCO-$\star$ and annealed VIMCO-GM, respectively, achieve $\widehat{k}\leq 0.7$, compared with only 24\% for ELBO and 16--18\% for the fixed VIMCO variants. Iterative Damped PSIS further improves reliability across all methods. The corresponding effective sample sizes of the importance weights are reported in Section~\ref{sec:supp-applications}. 

Computational differences are negligible at this dimension, with mean optimisation times ranging from 0.91 to 1.38 seconds across the four VIMCO variants, compared with 1.13 seconds for ELBO. 
Thus, although the VIMCO methods require more computation per
iteration to construct their gradient estimators, despite using
the same total Monte Carlo budget of $SN=300$ as ELBO, their
overall computational costs remain comparable to those of ELBO.
The corresponding selected or terminal values of $\alpha$ and optimisation iteration counts are reported in Table~\ref{tab:supp-alpha-iter-app}.

\FloatBarrier

\subsection{Riboflavin data}
\label{sec:application-riboflavin}

The Riboflavin dataset is a challenging high-dimensional regression problem with $n=71$ observations and $p=4088$ gene-expression predictors, where the response is the logarithm of the riboflavin production rate \citep{buhlmann2014high}. The predictors exhibit substantial correlation \citep{bar2020scalable}. Motivated by the sparse representations reported in \citet{bar2020scalable} and \citet{chung2026symmetric}, we specify $\pi\sim\operatorname{Beta}(0.00147,1)$, corresponding to a prior expected model size of approximately six. Since an exact posterior reference is unavailable, we evaluate the variational approximations through their out-of-sample predictive performance. For each method, we construct the posterior predictive distribution by drawing inclusion indicators from its variational approximation and averaging the corresponding model-specific predictive distributions, with the regression coefficients and residual variance integrated out analytically. We consider 50 repeated train--test partitions, each with 50 training and 21 test observations, and assess predictive RMSE and mean log predictive density (MLPD) on the test data.

\begin{figure}[h]
\centering
\includegraphics[width=\linewidth]{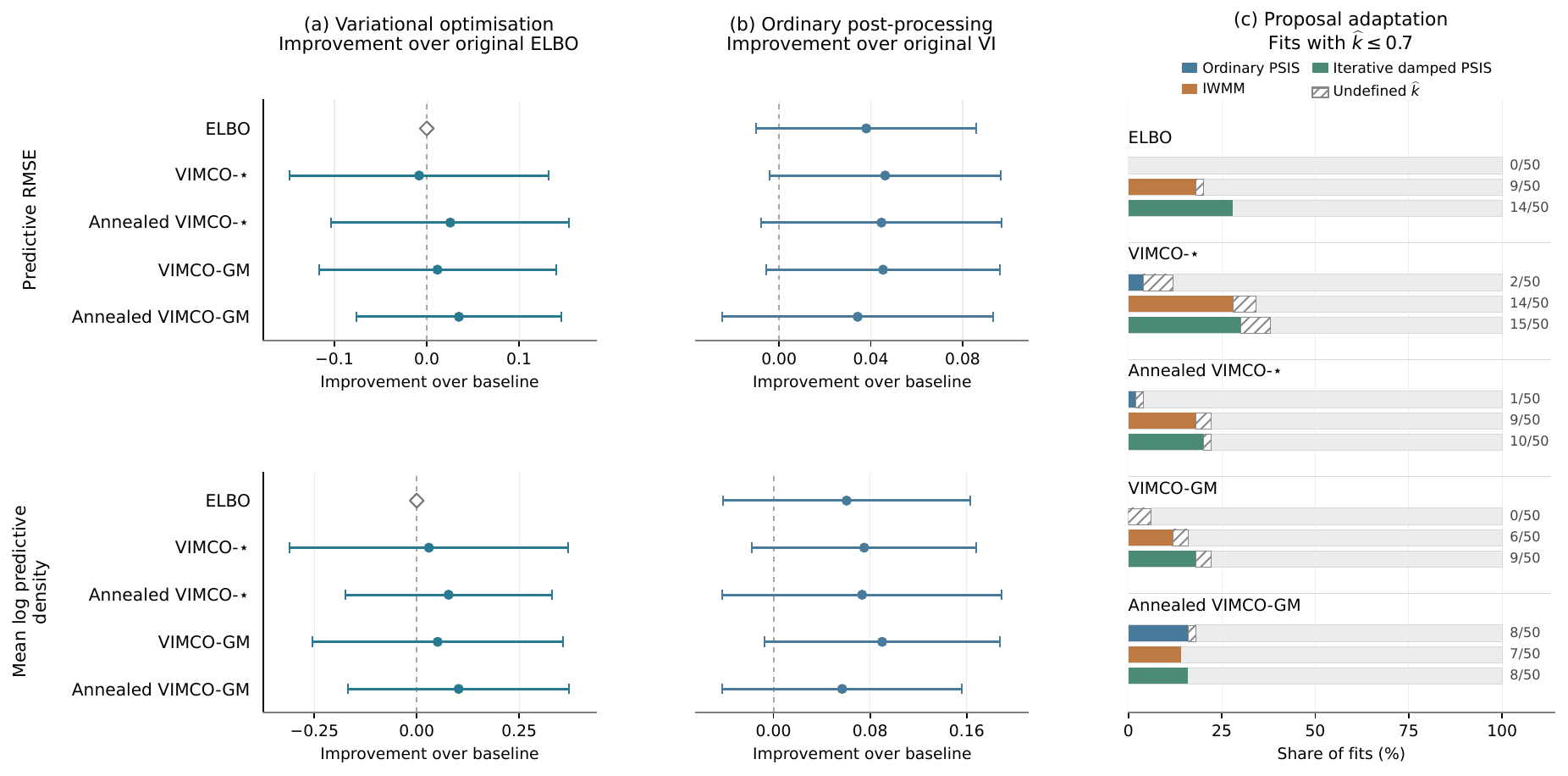}
\caption{Predictive accuracy and importance-sampling reliability in the Riboflavin application with $p=4088$ across 50 train--test splits. Panels (a) and (b) show mean paired improvements in predictive RMSE and mean log predictive density relative to original ELBO and the corresponding original variational approximation, respectively; positive values indicate improvements. Error bars represent one standard deviation across $50$ overlapping splits and are not confidence intervals. Panel (c) reports the percentage of all 50 fits with defined $\widehat{k}\leq0.7$ under ordinary PSIS, IWMM and iterative damped PSIS, and the remaining percentage has defined $\widehat{k}>0.7$.}
\label{fig:vimco-abc-riboflavin}
\end{figure}

Figure~\ref{fig:vimco-abc-riboflavin} presents the results on predictive performance and importance-sampling reliability.
Panel (a) shows that the annealed VIMCO variants yield improvements in mean predictive RMSE and MLPD relative to ELBO, with some variability across train--test splits, while the gains from the fixed-$\alpha$ variants are less pronounced. Panel (b) shows that ordinary PSIS further improves mean predictive performance across all methods, although these gains should be interpreted cautiously in light of the importance-sampling reliability diagnostics in panel (c). In particular, only 0--8 of the 50 fits achieve $\widehat{k}\leq 0.7$ under ordinary PSIS, and reliability remains poor even after proposal adaptation. Nevertheless, the improvements in predictive performance are encouraging given the substantial dimensionality of the problem. The corresponding effective sample sizes of the importance weights are reported in Section~\ref{sec:supp-applications}. The ESS-based annealing procedure terminates at a substantially higher value of $\alpha$ than in the body fat application, consistent with more severe importance-weight degeneracy in this high-dimensional setting; see Table~\ref{tab:supp-alpha-iter-app}.

The computational advantages of VIMCO are more pronounced in this high-dimensional application compared to that in Section~\ref{sec:application-bodyfat}. Mean optimisation time decreases from 102.3 seconds for ELBO to 45.0 and 67.9 seconds for VIMCO-$\star$ and annealed VIMCO-$\star$, respectively, and to 27.3 and 39.8 seconds for VIMCO-GM and annealed VIMCO-GM. 
These correspond to speed-ups of approximately 1.5--3.7 times relative to ELBO, showing that the improvements in mean predictive performance are accompanied by lower computational costs. The corresponding optimisation iteration counts are reported in Table~\ref{tab:supp-alpha-iter-app}.

\FloatBarrier

\section{Conclusion and future research}\label{sec:Conclusions}

We propose a collapsed importance-weighted variational framework for Bayesian variable selection that retains a simple, computationally convenient factorised approximation while improving posterior inclusion probability estimation. Our theoretical results establish a finite-sample bound on PIP approximation errors and show that, under suitable conditions, the optimised PIPs converge to their exact posterior counterparts as the number of importance samples increases and the R\'enyi parameter decreases appropriately. Importantly, this convergence does not require posterior independence among inclusion indicators, showing that the poor PIP accuracy under correlated predictors previously attributed to factorised variational approximations need not persist when optimising a tighter importance-weighted lower bound. We further establish that analytical marginalisation tightens the variational objective and PIP error bound, while reducing importance-weight variance and improving the asymptotic efficiency of importance-sampling post-processing. Our simulation experiments and empirical applications provide complementary evidence, demonstrating improvements in PIP estimation accuracy, variable selection and predictive performance relative to the standard ELBO objective under the same collapsed formulation. We also demonstrate more accurate PIP estimation relative to two existing variational approaches in the literature. Our results further suggest that adaptive post-processing can improve importance-sampling reliability, although the benefits depend on the quality of the initial variational approximation.

A natural extension is to consider more elaborate models where analytical marginalisation is possible only conditional on latent variables introduced through data augmentation. Examples include ordinal and continuous--ordinal regression using probit augmentation \citep{albert1993bayesian,kwon2007biomarkers,lee2026pcgs}, and logistic regression using P\'olya--gamma augmentation \citep{polson2013bayesian,bogoni2025bayesian}. Such extensions would retain the benefits of analytical marginalisation while requiring variational inference over both the inclusion indicators and augmentation variables. Extending our methodology and theoretical guarantees to these partially collapsed settings is a promising direction for future research. Another possible direction is to investigate more flexible variational families that capture dependence among inclusion indicators, potentially improving the reliability of importance-sampling post-processing.

\bibliographystyle{apalike}
\bibliography{ref_vimco}

\clearpage

\renewcommand{\thesection}{S\arabic{section}}

\renewcommand{\thesubsection}{\thesection.\arabic{subsection}}

\renewcommand{\theequation}{S\arabic{section}.\arabic{equation}}

\setcounter{section}{0}
\setcounter{equation}{0}
\setcounter{lemma}{0}
\setcounter{remark}{0}
\setcounter{proposition}{0}
\setcounter{corollary}{0}
\setcounter{figure}{0}
\setcounter{table}{0}
\setcounter{assumption}{0}

\renewcommand{\thefigure}{S\arabic{figure}}
\renewcommand{\thetable}{S\arabic{table}}
\renewcommand{\thelemma}{S\arabic{lemma}}
\renewcommand{\theremark}{S\arabic{remark}}
\renewcommand{\theproposition}{S\arabic{proposition}}
\renewcommand{\theassumption}{S\arabic{assumption}}
\renewcommand{\thecorollary}{S\arabic{corollary}}

\setcounter{page}{1}
\renewcommand{\thepage}{S\arabic{page}}

\section{Additional material}

\subsection{Implementation of importance-sampling post-processing}
\label{sec:supp-postprocessing}

We use the fitted product-Bernoulli variational distribution
$q_0(\vgamma)=q_{\widehat{\vlambda}}(\vgamma)$ as the initial
proposal for each post-processing procedure. For a current
proposal
\[
q_t(\vgamma)
=
\prod_{j=1}^p
(\rho_j^{(t)})^{\gamma_j}
(1-\rho_j^{(t)})^{1-\gamma_j},
\]
we independently generate $M$ model configurations
$\vgamma_t^{(1)},\ldots,\vgamma_t^{(M)}$ and calculate
the log importance weights
\[
a_{m,t}
=
\log p(\mathbf y,\vgamma_t^{(m)})
-
\log q_t(\vgamma_t^{(m)}),
\qquad m=1,\ldots,M.
\]
The unknown marginal likelihood cancels upon normalisation.
Pareto-smoothed importance sampling (PSIS)
\citep{vehtari2024pareto} regularises the upper tail of
the importance weights and yields normalised weights
$\widetilde w_{m,t}^{\,\mathrm{PSIS}}$ and the Pareto
shape diagnostic $\widehat k_t$. The corresponding
PSIS-weighted inclusion probabilities are
\begin{equation}
\widehat\rho_{j,t}^{\,\mathrm{PSIS}}
=
\sum_{m=1}^{M}
\widetilde w_{m,t}^{\,\mathrm{PSIS}}
\gamma_{j,t}^{(m)},
\qquad j=1,\ldots,p.
\label{eq:supp-psis-pip}
\end{equation}
Ordinary PSIS uses these estimates without modifying
the fitted proposal $q_0$.

In discrete importance-weighted moment matching (IWMM),
we update the Bernoulli probabilities by matching the
PSIS-weighted first moments,
\begin{equation}
\rho_j^{(t+1)}
=
\widehat\rho_{j,t}^{\,\mathrm{PSIS}},
\qquad j=1,\ldots,p.
\label{eq:supp-iwmm-update}
\end{equation}
Equivalently, this update maximises the weighted log
likelihood
$\sum_{m=1}^M \widetilde w_{m,t}^{\,\mathrm{PSIS}}
\log q_{\boldsymbol\rho}(\vgamma_t^{(m)})$
over product-Bernoulli proposals, corresponding to the
parameter update of the cross-entropy method for independent
Bernoulli families \citep{de2005tutorial}. Unlike the
original IWMM procedure, which applies affine
transformations to continuous draws
\citep{paananen2021implicitly}, our discrete adaptation
updates the Bernoulli parameters and generates new model
configurations. The procedure is therefore a parametric
adaptive importance-sampling scheme \citep{bugallo2017adaptive}.
To preserve full support on $\{0,1\}^p$, updated
probabilities are truncated to $[\delta,1-\delta]$,
where $\delta>0$ is a small numerical tolerance.

Because full moment matching may produce large changes
when the weighted inclusion probabilities are noisy,
iterative damped PSIS instead considers
\begin{equation}
\rho_{j,\omega}^{(t+1)}
=
(1-\omega)\rho_j^{(t)}
+
\omega\widehat\rho_{j,t}^{\,\mathrm{PSIS}},
\qquad \omega\in[0,1].
\label{eq:supp-damped-update}
\end{equation}
Here $\omega=0$ leaves the proposal unchanged,
$\omega=1$ gives the full moment-matching update, and
intermediate values provide more conservative updates.
This corresponds to the smoothed updating rule of the
cross-entropy method \citep[Remark~2.3]{de2005tutorial}.
At each iteration, we consider
$\omega\in\{0,0.25,0.50,0.75,1\}$ and evaluate the
corresponding candidate proposals using the Pareto
shape diagnostic obtained from new importance samples.
The candidate with the smallest $\widehat k$ is retained
only if it improves on the current proposal.
Support-preserving truncation is applied when needed.

Both adaptive procedures start independently from the
same fitted proposal $q_0$. An update is accepted only
when it reduces $\widehat k$, and adaptation terminates
when $\widehat k\leq0.7$, no further reduction is obtained,
or 20 iterations have been completed. An improvement may
be retained even if the resulting $\widehat k$ remains
above 0.7; the threshold is a stopping criterion rather
than an acceptance requirement. After adaptation, we draw model configurations from the
final retained proposal and estimate PIPs using their
corresponding normalised PSIS weights. The same weights
are used to combine model-specific conditional distributions
when evaluating latent-signal uncertainty and posterior
predictive performance. The Pareto shape diagnostic
$\widehat k$ assesses importance-weight reliability rather
than posterior approximation accuracy, so a lower $\widehat k$ does not
necessarily imply more accurate posterior estimates.

\subsection{Estimation of the control variates}
\label{sec:appendix:Estimating the Control Variates}

This section details the control-variate constructions used for VIMCO-$\star$ and VIMCO-GM in Section~\ref{sec:stochastic-optimisation}, following \citet{daudel2026importance}. We also describe the variance-reducing baseline used for the collapsed ELBO comparator. Throughout, write $u_{si}=w_{si}^{(\alpha)}$ for the tempered importance weights defined in Section~\ref{sec:stochastic-optimisation} and $s_{sij}=s_j(\vgamma_i^{(s)})=\gamma_{ij}^{(s)}-\rho_j$ for the corresponding score components.

For the VIMCO estimators, the baseline in \eqref{eq:VIMCO-gradient} is constructed as
\begin{equation}
 b_{-i,s,j}^{(\alpha)}
 =\frac{1}{1-\alpha}
 \log\left(
 \frac{\sum_{r\ne i}u_{sr}+c_{-i,s,j}^{(\alpha)}}{N}
 \right),
 \label{eq:CV-baseline}
\end{equation}
where $c_{-i,s,j}^{(\alpha)}$ is a nonnegative term calculated without using the current draw $\vgamma_i^{(s)}$.  This independence is essential for preserving the unbiasedness of the gradient estimator. To see this, let $\mathcal F_{-i,s}$ contain the optimisation history and all current draws other than $\vgamma_i^{(s)}$. Conditional on this information, $b_{-i,s,j}^{(\alpha)}$ is fixed and the omitted draw has a zero-mean score. Hence
\[
 \E\!\left[s_{sij}b_{-i,s,j}^{(\alpha)}\mid\mathcal F_{-i,s}\right]
 =b_{-i,s,j}^{(\alpha)}\E[s_{sij}\mid\mathcal F_{-i,s}]
 =0.
\]
The baseline therefore changes the variance of the gradient estimator without changing its expectation. This applies both to a leave-one-out construction using the other draws in the current group and to a baseline estimated from an earlier optimisation batch.

For VIMCO-$\star$, the asymptotically variance-minimising limiting term in \citet[Theorem~5]{daudel2026importance}, for fixed $0<\alpha<1$, has population value
\begin{equation}
 c_j^{(\alpha,\star)}
 =\alpha\,
 \frac{
 \E_{q_{\vlambda}}\!\left[w_{\vlambda}(\vgamma;\mathbf y)^{1-\alpha}s_j(\vgamma)^2\right]
 -\displaystyle\frac{
 \E_{q_{\vlambda}}\!\left[w_{\vlambda}(\vgamma;\mathbf y)^{1-\alpha}s_j(\vgamma)\right]^2
 }{
 \E_{q_{\vlambda}}\!\left[w_{\vlambda}(\vgamma;\mathbf y)^{1-\alpha}\right]
 }
 }{
 \mathbb V_{q_{\vlambda}}[s_j(\vgamma)]
 }.
 \label{eq:CV-star-population}
\end{equation}
The numerator is nonnegative by Cauchy--Schwarz, and for our product-Bernoulli family the denominator is $\rho_j(1-\rho_j)>0$ for finite $\lambda_j$. The required expectations are not available in closed form, so the control variate is estimated from Monte Carlo moments. For a moment-estimation batch of size $M_0$, let $u_r$ and $s_{rj}$ denote its tempered weights and score components, and define
\begin{equation}
 A_{k,\ell,j}^{(\alpha)}
 =\frac{1}{M_0}\sum_{r=1}^{M_0}u_r^k s_{rj}^{\ell}.
 \label{eq:CV-star-moments}
\end{equation}
The corresponding empirical control-variate term is
\begin{equation}
 \widehat c_j^{(\alpha,\star)}
 =\alpha\,
 \frac{
 A_{1,2,j}^{(\alpha)}
 -\bigl(A_{1,1,j}^{(\alpha)}\bigr)^2/A_{1,0,j}^{(\alpha)}
 }{
 A_{0,2,j}^{(\alpha)}-\bigl(A_{0,1,j}^{(\alpha)}\bigr)^2
 },
 \label{eq:CV-star-MC}
\end{equation}
when the empirical score variance in the denominator is positive. Although \citet{daudel2026importance} also give a leave-one-out estimator based on the other $N-1$ draws, our implementation estimates these moments from the preceding optimisation batch and holds the resulting value fixed for the current gradient batch. Thus the same lagged, coordinate-specific estimate is used for $c_{-i,s,j}^{(\alpha)}$ across the current draws, avoiding repeated leave-one-out moment calculations while retaining the required conditional unbiasedness. At $\alpha=0$, the VIMCO-$\star$ term is set to zero. Under annealing, the lagged estimate need not be exactly optimal at the current parameter values; its independence from the current draws is what preserves unbiasedness.

For VIMCO-GM, the excluded tempered weight is instead replaced by the geometric mean of the other weights in the same group \citep{mnih2016variational,daudel2026importance}:
\begin{equation}
 c_{-i,s,j}^{(\alpha,\mathrm{GM})}
 =\exp\left(
 \frac{1}{N-1}\sum_{r\ne i}\log u_{sr}
 \right).
 \label{eq:CV-GM}
\end{equation}
This term does not depend on $j$ or the omitted draw, and it requires only the weights already generated for estimating the VR-IWAE objective. It therefore avoids the additional moment-estimation step of VIMCO-$\star$. Substituting either construction into \eqref{eq:CV-baseline} and then \eqref{eq:VIMCO-gradient} gives the two gradient estimators used in our experiments. Both target the same VR-IWAE gradient; only their control variates differ. The favourable fixed-$\alpha$ variance and signal-to-noise orders stated in Section~\ref{sec:stochastic-optimisation} are for $0<\alpha<1$; in particular, at $\alpha=0$ the geometric-mean estimator reduces to the original VIMCO baseline in \citet{mnih2016variational}.

For a comparable implementation, the ELBO baseline also operates on the same collapsed target, with $h_{\vlambda}(\vgamma)=\log p(\mathbf y,\vgamma)-\log q_{\vlambda}(\vgamma)$. From $M_{\mathrm E}$ independent draws $\vgamma_1,\ldots,\vgamma_{M_{\mathrm E}}\sim q_{\vlambda}$, its score-function gradient estimate is
\begin{equation}
 \widehat g_j^{\mathrm{ELBO}}
 =\frac{1}{M_{\mathrm E}}\sum_{r=1}^{M_{\mathrm E}}
 s_j(\vgamma_r)\{h_{\vlambda}(\vgamma_r)-d_j\}.
 \label{eq:CV-ELBO-gradient}
\end{equation}
Because $\E_{q_{\vlambda}}[s_j(\vgamma)]=0$, any baseline $d_j$ fixed independently of the current batch leaves the expected gradient unchanged. The variance-minimising scalar baseline is
\begin{equation}
 d_j^{\star}
 =\frac{
 \operatorname{Cov}_{q_{\vlambda}}\!\left(s_j(\vgamma)h_{\vlambda}(\vgamma),s_j(\vgamma)\right)
 }{
 \mathbb V_{q_{\vlambda}}[s_j(\vgamma)]
 }.
 \label{eq:CV-ELBO-optimal}
\end{equation}
As with VIMCO-$\star$, this quantity is estimated from one Monte Carlo batch and applied to the next, so that it is fixed conditional on the optimisation history when the new gradient samples are drawn. The optimisation and tuning settings shared by the methods are given in Section~\ref{sec:supp-implementation}.

\subsection{Implementation details}
\label{sec:supp-implementation}

All variational methods except \texttt{varbvs} \citep{carbonetto2012scalable} and approximated collapsed variational Bayes (ACVB; \citealp{you2023approximate}) are fitted using Adam
\citep{KingmaBa2015Adam} with a common per-iteration budget of 300 Monte
Carlo samples. We use $S=1$ and $N=300$ for the VIMCO-based methods, and
300 samples for the ELBO. For each dataset and method, the Adam learning
rate is selected from $\{0.01,0.05,0.10,0.20,0.30\}$ and, for
fixed-$\alpha$ VIMCO, $\alpha$ is additionally selected from
$\{0.2,0.5,0.7,0.9\}$. For annealed VIMCO, $\alpha$ is initialised at
$0.99$ and decreased by $0.01$ whenever the tempered-weight ESS in
\eqref{eq:annealing-ESS} reaches at least $80\%$ of $SN$ (i.e., 240),
following Section~\ref{sec:stochastic-optimisation}.

Candidate configurations are compared using an independently evaluated
standard ELBO based on 2000 draws. Configurations within two combined
Monte Carlo standard errors of the largest evaluated ELBO are treated
as indistinguishable, and the one with the smallest Monte Carlo
standard error is selected. Optimisation stops when the training
objective has not improved by more than $10^{-4}$ over 200 consecutive
iterations, or after 3000 iterations; the iterate with the largest
training objective is retained. For annealed VIMCO, a decrease in
$\alpha$ resets this stopping rule. Post-processing is applied only
to the selected fit, using $M=5000$ importance-sampling draws.

\subsection{Additional results for the experiments}\label{sec:additional-experiment-results}
Table~\ref{tab:postprocess-ess} reports the median effective sample size (ESS) of the importance weights obtained from each post-processing procedure. The VIMCO methods generally yield higher ESS than ELBO, with particularly pronounced differences at $p=1000$ and $p=5000$. Proposal adaptation often further increases ESS relative to ordinary PSIS, although the improvements vary across methods and settings.

\begin{table}[h]
\centering
\caption{Median effective sample size (ESS) of the importance weights obtained from each post-processing procedure in the simulation experiments in Section~\ref{sec:linear-regression-experiments}. Each entry is the median across 50 independently simulated datasets, using 5,000 importance draws per fit.}
\label{tab:postprocess-ess}
\small
\setlength{\tabcolsep}{5pt}
\begin{tabular}{llrrr}
\toprule
Setting & Variational method & Ordinary PSIS & IWMM & Iterative damped PSIS \\
\midrule
$p=30$ & ELBO & 830.4 & 2048.1 & 2010.3 \\
 & VIMCO-$\star$ & 1881.7 & 2791.4 & 2608.5 \\
 & Annealed VIMCO-$\star$ & 2648.9 & 2899.7 & 2887.5 \\
 & VIMCO-GM & 2559.3 & 2773.8 & 2632.8 \\
 & Annealed VIMCO-GM & 2667.7 & 2667.7 & 2797.8 \\
\addlinespace
$p=1000$ & ELBO & 10.8 & 69.1 & 73.9 \\
 & VIMCO-$\star$ & 137.3 & 219.6 & 140.9 \\
 & Annealed VIMCO-$\star$ & 218.0 & 316.1 & 325.0 \\
 & VIMCO-GM & 121.3 & 188.6 & 198.5 \\
 & Annealed VIMCO-GM & 229.8 & 238.9 & 275.5 \\
\addlinespace
$p=5000$ & ELBO & 2.5 & 5.1 & 88.6 \\
 & VIMCO-$\star$ & 364.8 & 664.1 & 608.0 \\
 & Annealed VIMCO-$\star$ & 977.7 & 1042.2 & 1210.4 \\
 & VIMCO-GM & 546.9 & 805.2 & 805.2 \\
 & Annealed VIMCO-GM & 900.6 & 928.6 & 928.6 \\
\bottomrule
\end{tabular}
\end{table}

\begin{figure}[h]
\centering
\includegraphics[width=\linewidth]{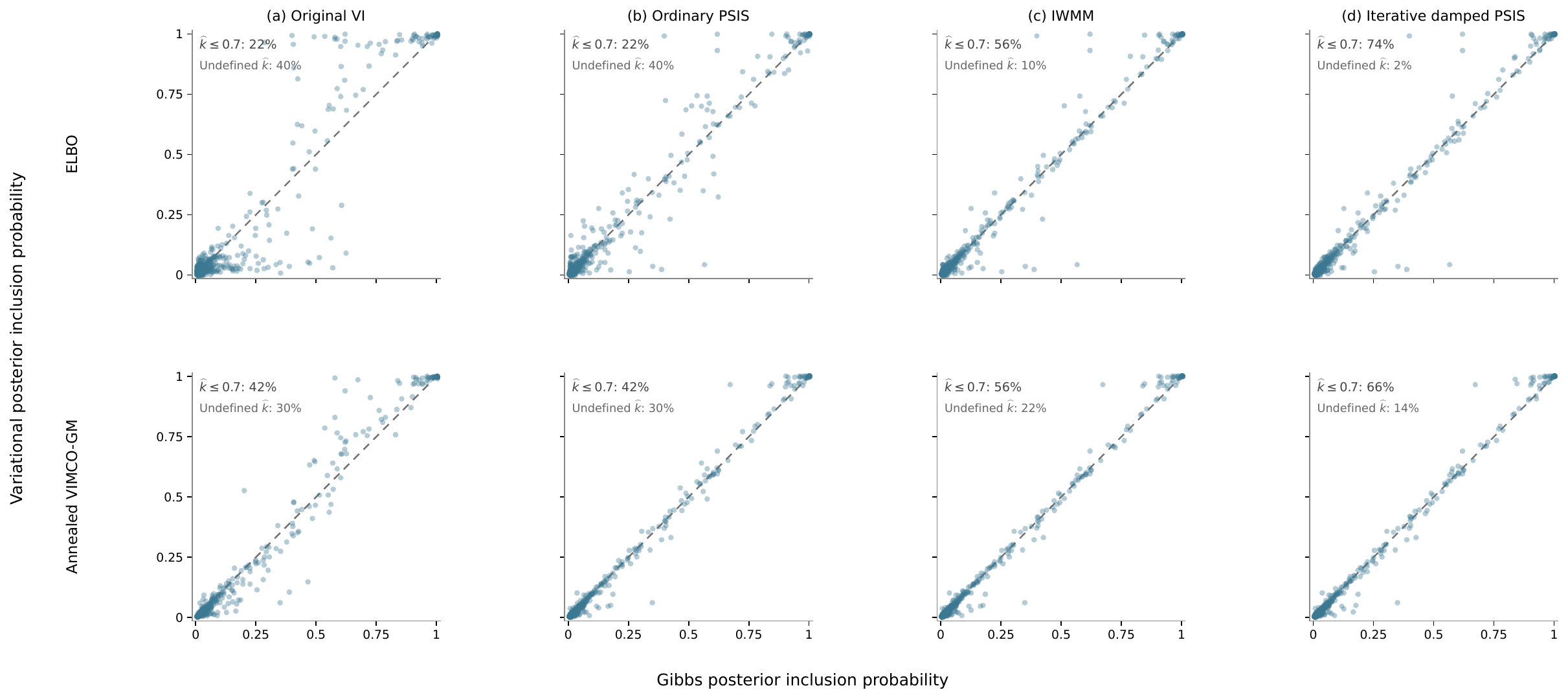}
\caption{Predictor-level posterior inclusion probabilities (PIPs) for ELBO (top row) and annealed VIMCO-GM (bottom row) at $p=30$, before and after post-processing using ordinary PSIS, IWMM and iterative damped PSIS. Each panel includes all 50 simulated datasets and all 30 predictors per dataset, regardless of the Pareto diagnostic. The percentage annotated within each panel is the share of all 50 fits with defined $\widehat{k}\leq0.7$; undefined diagnostics are identified separately where present. The original VI and ordinary PSIS panels share a diagnostic because ordinary PSIS leaves the proposal unchanged. The dashed diagonal denotes agreement with the corresponding collapsed Gibbs reference.}
\label{fig:vimco-pips-p30-postprocessing}
\end{figure}

\begin{figure}[h]
\centering
\includegraphics[width=\linewidth]{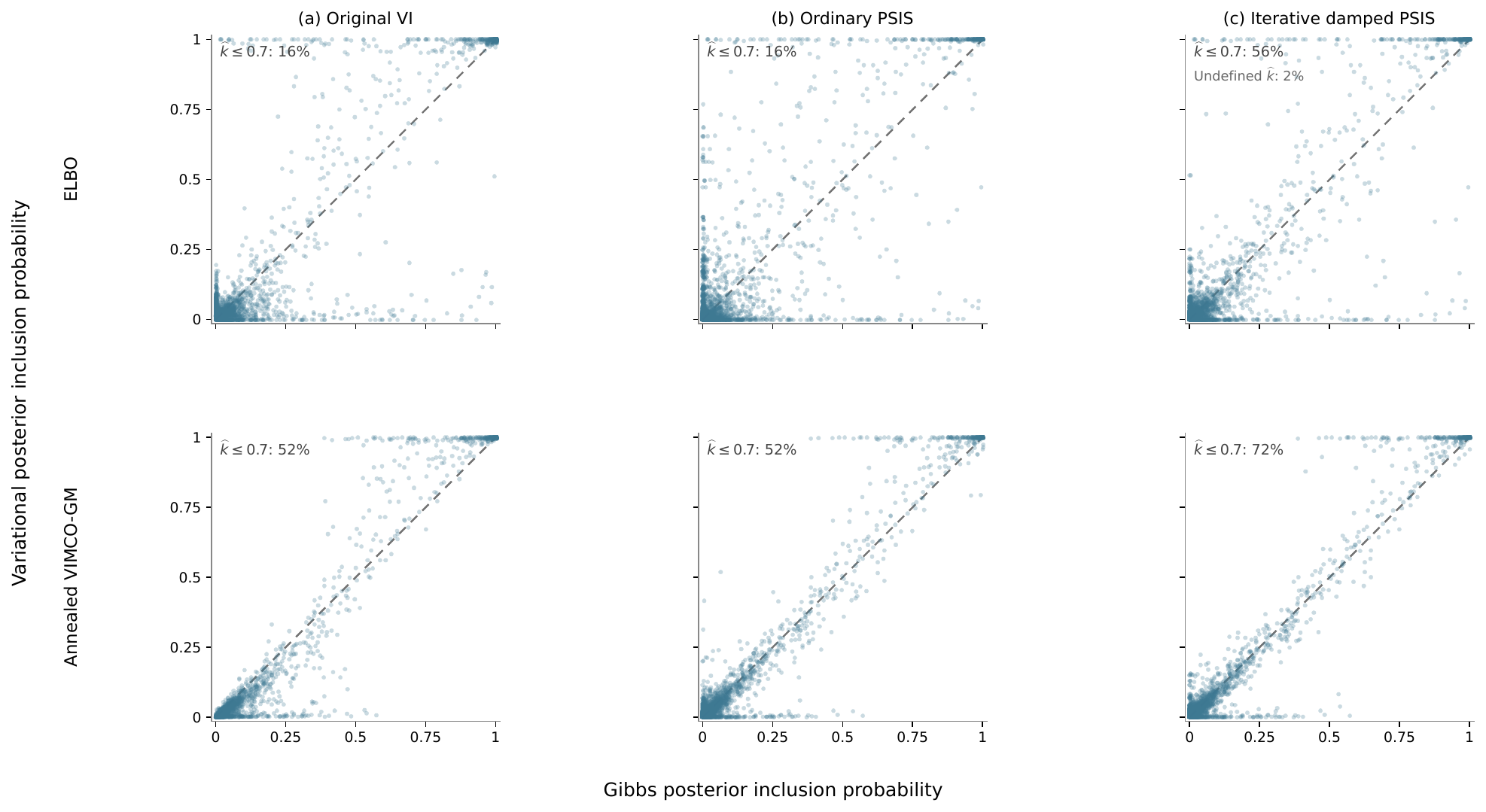}
\caption{Predictor-level posterior inclusion probabilities (PIPs) for ELBO (top row) and annealed VIMCO-GM (bottom row) at $p=1000$, before and after post-processing using ordinary PSIS and iterative damped PSIS. The panel definitions and plotting details are as in Figure~\ref{fig:vimco-pips-p30-postprocessing}.}
\label{fig:vimco-pips-p1000-postprocessing}
\end{figure}

Figures~\ref{fig:vimco-pips-p30-postprocessing} and~\ref{fig:vimco-pips-p1000-postprocessing} provide a predictor-level illustration of the effects of importance-sampling post-processing for ELBO and annealed VIMCO-GM, for $p=30$ and $p=1000$, respectively. We focus on annealed VIMCO-GM as a representative VIMCO method because it consistently achieves high PIP accuracy across both dimensions. At $p=30$, ordinary PSIS brings the estimated PIPs closer to the Gibbs reference for both methods, while proposal adaptation further reduces the discrepancies, particularly for ELBO. Iterative damped PSIS yields especially close agreement with the Gibbs reference, although some deviations remain. The improvements are also reflected in the importance-sampling diagnostics: the proportion of fits satisfying $\widehat{k}\leq0.7$ increases from 22\% under ordinary PSIS to 74\% under iterative damped PSIS for ELBO, and from 42\% to 66\% for annealed VIMCO-GM. At $p=1000$, the differences between the two variational methods become more pronounced. The original annealed VIMCO-GM approximation exhibits substantially closer agreement with the Gibbs reference than ELBO, and this advantage is largely maintained after post-processing. Ordinary PSIS does not resolve the substantial discrepancies exhibited by ELBO, whereas iterative damped PSIS improves the agreement for some predictors but does not eliminate the larger deviations. The deterioration in PIP accuracy for ELBO under ordinary PSIS is unsurprising given its unfavourable importance-sampling diagnostics at this dimension. Nevertheless, proposal adaptation improves the importance-sampling diagnostics for both methods, increasing the proportion of fits satisfying $\widehat{k}\leq0.7$ from 16\% to 56\% for ELBO and from 52\% to 72\% for annealed VIMCO-GM. 

\FloatBarrier

Table~\ref{tab:supp-alpha-iter-sim} summarises the optimisation behaviour of
the selected fits in the simulation experiments. With fixed $\alpha$, the
importance weights need to remain sufficiently stable from initialisation,
when the variational approximation is still far from the posterior, so the
selected values lie near the upper end of the grid; the evaluated-ELBO
selection criterion, which measures the reverse KL divergence, may also
favour the more concentrated fits obtained with larger $\alpha$.
Annealing mitigates this trade-off by starting near $\alpha=1$ and
decreasing $\alpha$ only once the tempered weights are sufficiently
balanced, so its terminal value indicates how far the objective can be
tightened in practice. This limit depends on the setting: annealing
reaches about 0.19 at $p=30$ but tends to terminate at higher values
(0.36--0.56) in higher dimensions, consistent with more severe
importance-weight degeneracy
\citep{chatterjee2018sample, vehtari2024pareto}.
This may also explain why the additional benefit of annealing is smaller
at $p=1000$. At $p=1000$ and $p=5000$, the VIMCO variants terminate
in 36--58\% fewer iterations than ELBO, consistent with their lower
runtimes in Table~\ref{tab:simulation-runtime}.

\begin{table}[h]
\centering
\caption{Mean values of $\alpha$ and optimisation iteration counts across 50 independently simulated datasets for each setting. For fixed-$\alpha$ methods, $\alpha$ is selected by tuning; for annealed methods, its terminal value is reported.}
\label{tab:supp-alpha-iter-sim}
\small
\setlength{\tabcolsep}{5pt}
\begin{tabular}{llrr}
\toprule
Setting & Variational method & $\alpha$ & Iterations \\
\midrule
$p=30$ & ELBO & -- & 411.5 \\
 & VIMCO-$\star$ & 0.868 & 596.3 \\
 & Annealed VIMCO-$\star$ & 0.190 & 536.5 \\
 & VIMCO-GM & 0.860 & 620.2 \\
 & Annealed VIMCO-GM & 0.194 & 514.0 \\
\addlinespace
$p=1000$ & ELBO & -- & 1852.9 \\
 & VIMCO-$\star$ & 0.848 & 1020.9 \\
 & Annealed VIMCO-$\star$ & 0.561 & 815.4 \\
 & VIMCO-GM & 0.856 & 889.8 \\
 & Annealed VIMCO-GM & 0.559 & 781.0 \\
\addlinespace
$p=5000$ & ELBO & -- & 2351.0 \\
 & VIMCO-$\star$ & 0.864 & 1514.4 \\
 & Annealed VIMCO-$\star$ & 0.360 & 1358.6 \\
 & VIMCO-GM & 0.872 & 1325.2 \\
 & Annealed VIMCO-GM & 0.402 & 1086.6 \\
\bottomrule
\end{tabular}
\end{table}

\FloatBarrier

\subsection{Additional results for the applications}\label{sec:supp-applications}

Table~\ref{tab:postprocess-ess-applications} reports the median effective sample sizes of the importance weights for the two empirical applications. For body fat, the ESS values are relatively high across all methods, with proposal adaptation generally providing further improvements. In contrast, ESS values are substantially lower for Riboflavin, reflecting the challenges of importance-sampling post-processing in this high-dimensional setting. Although proposal adaptation generally improves ESS, the values remain low across all methods, consistent with the importance-sampling reliability difficulties discussed in Section~\ref{sec:application-riboflavin}.

\begin{table}[h]
\centering
\caption{Median effective sample size (ESS) of the importance weights obtained from each post-processing procedure in the empirical applications in Section~\ref{sec:Applications}. Each entry is the median across 50 fits, using 5,000 importance draws per fit. For body fat, the fits correspond to repeated optimisations of the same data; for riboflavin, they correspond to repeated train--test splits. }
\label{tab:postprocess-ess-applications}
\small
\setlength{\tabcolsep}{5pt}
\begin{tabular}{llrrr}
\toprule
Application & Variational method & Ordinary PSIS & IWMM & Iterative damped PSIS \\
\midrule
Body fat & ELBO & 2062.2 & 2574.2 & 2566.5 \\
 & VIMCO-$\star$ & 1464.9 & 2240.2 & 2530.3 \\
 & Annealed VIMCO-$\star$ & 2322.4 & 2588.9 & 2627.8 \\
 & VIMCO-GM & 1490.2 & 2588.2 & 2540.5 \\
 & Annealed VIMCO-GM & 2227.8 & 2471.1 & 2553.6 \\
\addlinespace
Riboflavin & ELBO & 14.8 & 20.6 & 24.1 \\
 & VIMCO-$\star$ & 32.0 & 44.8 & 46.7 \\
 & Annealed VIMCO-$\star$ & 26.0 & 32.1 & 29.4 \\
 & VIMCO-GM & 33.2 & 24.9 & 30.0 \\
 & Annealed VIMCO-GM & 25.9 & 28.5 & 32.4 \\
\bottomrule
\end{tabular}
\end{table}

Table~\ref{tab:supp-alpha-iter-app} reports the mean selected or terminal
values of $\alpha$ and the mean optimisation iteration counts for the two
applications. The fixed-$\alpha$ variants again select values towards the
upper end of the grid. As at $p=30$, annealing reaches about 0.16--0.17
for the low-dimensional Body fat data, whereas for Riboflavin it
terminates near $\alpha\approx0.69$, only modestly below the selected
fixed values. This is consistent with more severe importance-weight
degeneracy in this high-dimensional setting with strongly correlated
predictors, which limits how far $\alpha$ can be decreased. This may
partly explain why the predictive improvements over ELBO in
Section~\ref{sec:application-riboflavin} are modest and variable across train--test
splits. Iteration counts are comparable across methods for Body fat,
whereas for Riboflavin the VIMCO variants terminate in approximately
44--62\% fewer iterations than ELBO, consistent with their
1.5--3.7-fold speed-ups in optimisation time.

\begin{table}[h]
\centering
\caption{Mean selected or terminal values of $\alpha$ and optimisation
iteration counts for the two empirical applications. Means are computed
across 50 optimisation runs on the same dataset for Body fat, and
across 50 train--test splits for Riboflavin. For fixed-$\alpha$
methods, $\alpha$ is selected by tuning; for annealed methods,
its terminal value is reported.}
\label{tab:supp-alpha-iter-app}
\small
\setlength{\tabcolsep}{5pt}
\begin{tabular}{llrr}
\toprule
Application & Variational method & $\alpha$ & Iterations \\
\midrule
Body fat & ELBO & -- & 591.1 \\
 & VIMCO-$\star$ & 0.782 & 588.9 \\
 & Annealed VIMCO-$\star$ & 0.160 & 685.2 \\
 & VIMCO-GM & 0.760 & 560.2 \\
 & Annealed VIMCO-GM & 0.171 & 654.9 \\
\addlinespace
Riboflavin & ELBO & -- & 2247.3 \\
 & VIMCO-$\star$ & 0.816 & 1192.9 \\
 & Annealed VIMCO-$\star$ & 0.694 & 1266.6 \\
 & VIMCO-GM & 0.860 & 861.3 \\
 & Annealed VIMCO-GM & 0.693 & 1080.3 \\
\bottomrule
\end{tabular}
\end{table}
\FloatBarrier

\section{Proofs}
\begin{proof}[Proof of Lemma~\ref{lem:VR-IWAE-PIP}]
Let $P=p(\vgamma|\mathbf y)$ and $Q=q_{\vlambda}(\vgamma)$, with $P(\gamma_j=1)=\Pr(\gamma_j=1|\mathbf y)$ and $Q(\gamma_j=1)=\rho_j$. We have
\[
|P(\gamma_j=1)-Q(\gamma_j=1)|
\leq
\operatorname{TV}(P,Q),
\]
where $\operatorname{TV}$ denotes the total variation distance. The generalised Pinsker inequality \citep[Theorem~31]{vanerven2014renyi} gives
\[
D_{1-\alpha}(P\|Q)
\geq
2(1-\alpha)\operatorname{TV}(P,Q)^2.
\]
Combining this inequality with \eqref{eq:VR-IWAE-renyi}, we obtain
\[
\begin{aligned}
\log p(\mathbf y)-\ell_N^{(\alpha)}(\vlambda)
&\geq
\frac{\alpha}{1-\alpha}D_{1-\alpha}(P\|Q)\\
&\geq
2\alpha\operatorname{TV}(P,Q)^2\\
&\geq
2\alpha|P(\gamma_j=1)-Q(\gamma_j=1)|^2.
\end{aligned}
\]
The result follows.
\end{proof}

\begin{proof}[Proof of Theorem~\ref{thm:PIP-convergence}]
The proof begins by showing that the forward
Kullback--Leibler (KL) divergence from the true
posterior to a product-Bernoulli distribution is minimised
by matching the exact posterior inclusion probabilities,
irrespective of dependence among the inclusion indicators.
It then proceeds in two steps. First, we establish a uniform
bound on the approximation error of forward KL by a
rescaled VR-IWAE gap, through (a) bounding the
approximation error of a rescaled variational
R\'enyi gap and (b) controlling the additional finite-$N$
approximation error introduced by VR-IWAE. Second,
we combine this uniform error bound with the global
optimality condition in Assumption~\ref{ass:PIP-convergence}(iii) to bound the difference between the forward KL divergence at the
VR-IWAE optimiser and its minimum, and then use
Pinsker's inequality to translate this difference
into a bound on the PIP approximation errors.

Write
\[
P(\vgamma)=p(\vgamma|\mathbf y),
\qquad
q_{\boldsymbol{\rho}}(\vgamma)
=
q_{\operatorname{logit}(\boldsymbol{\rho})}(\vgamma),
\]
and define
\[
\rho_j^\star=\Pr(\gamma_j=1|\mathbf y),
\qquad
\boldsymbol{\rho}^\star
=
(\rho_1^\star,\ldots,\rho_p^\star)^\top.
\]
By Assumption~\ref{ass:PIP-convergence}(i),
$0<\rho_j^\star<1$ for all $j=1,\ldots,p$.
Assumption~\ref{ass:PIP-convergence}(ii)
then ensures that
$\boldsymbol{\rho}^\star\in\mathcal R$. We first show that
$\boldsymbol{\rho}^\star$ is the unique minimiser
of the forward KL divergence
$\operatorname{KL}(P\|q_{\boldsymbol{\rho}})$
over $\mathcal R$. Our objective is to establish an approximation error
bound for the VR-IWAE maximiser
$\widehat{\boldsymbol{\rho}}_{N,\alpha}$
in \eqref{eq:VR-IWAE-global-maximiser}
relative to $\boldsymbol{\rho}^\star$,
which implies convergence to
$\boldsymbol{\rho}^\star$ under a suitable
choice of $\alpha$ as $N$ increases,
as stated in Corollary~1.

\medskip

\noindent\textbf{Forward KL minimisation
in the product-Bernoulli family.}
Consider the forward KL divergence
\begin{equation}\label{eq:forward_KL}
K(\boldsymbol{\rho})
=
\operatorname{KL}
\left(P\|q_{\boldsymbol{\rho}}\right)
=
\E_P\left[
\log\left(
\frac{P(\vgamma)}
{q_{\boldsymbol{\rho}}(\vgamma)}
\right)
\right].
\end{equation}
Using the product-Bernoulli structure in
\eqref{eq:variational-distribution}, we obtain
\[K(\boldsymbol{\rho})
=
\sum_{\vgamma}P(\vgamma)\log P(\vgamma)
-
\sum_{j=1}^p
\left[
\rho_j^\star\log\rho_j
+
(1-\rho_j^\star)\log(1-\rho_j)
\right].
\]
Evaluating this expression at
$\boldsymbol{\rho}=\boldsymbol{\rho}^\star$ gives
\[
K(\boldsymbol{\rho}^\star)
=
\sum_{\vgamma}P(\vgamma)\log P(\vgamma)
-
\sum_{j=1}^p
\left[
\rho_j^\star\log\rho_j^\star
+
(1-\rho_j^\star)\log(1-\rho_j^\star)
\right].
\]
Subtracting the two expressions yields
\begin{equation}
K(\boldsymbol{\rho})
-
K(\boldsymbol{\rho}^\star)
=
\sum_{j=1}^p
\operatorname{KL}
\left(
\operatorname{Bernoulli}(\rho_j^\star)
\|
\operatorname{Bernoulli}(\rho_j)
\right).
\label{eq:proof-forward-KL-projection}
\end{equation}
Since each term on the right-hand side is
nonnegative, with equality only when
$\rho_j=\rho_j^\star$, the unique minimiser of
$K(\boldsymbol{\rho})$ over $\mathcal R$ is
$\boldsymbol{\rho}^\star$. Importantly, this
identity does not require the posterior inclusion
indicators to be independent.

\medskip
\noindent\textbf{Step 1: Uniform error bounds for
variational approximations of forward KL.}\par
\noindent\textbf{(a) Uniform approximation error of
forward KL by a rescaled variational R\'enyi gap.}
Define the normalised importance weight
\begin{equation}\label{eq:norm_weights}
U_{\boldsymbol{\rho}}(\vgamma)
=
\frac{P(\vgamma)}
{q_{\boldsymbol{\rho}}(\vgamma)}.
\end{equation}
By Assumption~\ref{ass:PIP-convergence}(ii),
$q_{\boldsymbol{\rho}}(\vgamma)>0$ for all
$\vgamma\in\{0,1\}^p$ and
$\boldsymbol{\rho}\in\mathcal R$, so
$U_{\boldsymbol{\rho}}(\vgamma)$ is well defined.
Let $Z=p(\mathbf y)$, so that the unnormalised
importance weight satisfies
\[
w_{\vlambda}(\vgamma;\mathbf y)
=
ZU_{\boldsymbol{\rho}}(\vgamma),
\]
where $\vlambda=\operatorname{logit}(\boldsymbol{\rho})$.
Substituting this identity into the variational
R\'enyi bound in \eqref{eq:renyi-bound} yields
\[
\mathcal L_{\mathrm R}^{(\alpha)}
\left(\operatorname{logit}(\boldsymbol{\rho})\right)
=
\log Z
+
\frac{1}{1-\alpha}
\log
\E_{q_{\boldsymbol{\rho}}}
\left[
U_{\boldsymbol{\rho}}(\vgamma)^{1-\alpha}
\right].
\]

We want to relate the logarithmic expectation
above to the forward KL divergence
$K(\boldsymbol{\rho})$ in
\eqref{eq:forward_KL}. To this end, using
$q_{\boldsymbol{\rho}}(\vgamma)
U_{\boldsymbol{\rho}}(\vgamma)=P(\vgamma)$,
we obtain

\begin{align*}
\E_{q_{\boldsymbol{\rho}}}
\left[
U_{\boldsymbol{\rho}}(\vgamma)^{1-\alpha}
\right]
&=
\sum_{\vgamma}
q_{\boldsymbol{\rho}}(\vgamma)
U_{\boldsymbol{\rho}}(\vgamma)^{1-\alpha}\\
&=
\sum_{\vgamma}
P(\vgamma)
U_{\boldsymbol{\rho}}(\vgamma)^{-\alpha}\\
&=
\E_P
\left[
\exp\left(
-\alpha\log U_{\boldsymbol{\rho}}(\vgamma)
\right)
\right].
\end{align*}
Hence the variational
R\'enyi bound in \eqref{eq:renyi-bound}
can alternatively be expressed as
\begin{equation}\label{eq:VR_IWAE_alt}
\mathcal L_{\mathrm R}^{(\alpha)}
\left(\operatorname{logit}(\boldsymbol{\rho})\right)
=
\log Z
+
\frac{1}{1-\alpha}
\log
\E_P
\left[
\exp\left(
-\alpha\log U_{\boldsymbol{\rho}}(\vgamma)
\right)
\right].
\end{equation}

To establish the connection between the logarithmic
expectation and the forward KL, define
\[
V_{\boldsymbol{\rho}}(\vgamma)
=
\log U_{\boldsymbol{\rho}}(\vgamma),
\qquad
g_{\boldsymbol{\rho}}(t)
=
\log\E_P
\left[
\exp\left(-tV_{\boldsymbol{\rho}}(\vgamma)\right)
\right].
\]
We apply a Taylor expansion of
$g_{\boldsymbol{\rho}}(t)$ around $t=0$,
which corresponds to $\alpha=0$.
As in Theorem~\ref{thm:PIP-convergence},
we consider any fixed $\alpha_0\in(0,1)$
and evaluate the expansion for
$0<\alpha\leq\alpha_0$.
First, $g_{\boldsymbol{\rho}}(0)=0$.
Differentiating gives
\[
g_{\boldsymbol{\rho}}'(t)
=
-
\frac{
\E_P\left[
V_{\boldsymbol{\rho}}(\vgamma)
e^{-tV_{\boldsymbol{\rho}}(\vgamma)}
\right]
}{
\E_P\left[
e^{-tV_{\boldsymbol{\rho}}(\vgamma)}
\right]
}.
\]
Evaluating at $t=0$ and using
\eqref{eq:forward_KL}, we obtain
\[
g_{\boldsymbol{\rho}}'(0)
=
-\E_P[V_{\boldsymbol{\rho}}(\vgamma)]
=
-K(\boldsymbol{\rho}).
\]
Taylor's theorem therefore gives
\[
g_{\boldsymbol{\rho}}(\alpha)
=
-\alpha K(\boldsymbol{\rho})
+
R_{\boldsymbol{\rho}}(\alpha),
\]
equivalently,
\[
\log
\E_P
\left[
\exp\left(
-\alpha\log U_{\boldsymbol{\rho}}(\vgamma)
\right)
\right]
=
-\alpha K(\boldsymbol{\rho})
+
R_{\boldsymbol{\rho}}(\alpha),
\]
where $R_{\boldsymbol{\rho}}(\alpha)$ denotes
the second-order remainder. By Taylor's theorem,
the remainder satisfies
\[
|R_{\boldsymbol{\rho}}(\alpha)|
\leq
\frac{\alpha^2}{2}
\sup_{0\leq t\leq\alpha}
|g_{\boldsymbol{\rho}}''(t)|.
\]
To control the approximation error, it therefore
remains to bound the second derivative uniformly
over $\boldsymbol{\rho}\in\mathcal R$.

Differentiating again gives
\[
g_{\boldsymbol{\rho}}''(t)
=
\mathbb V_{P_{t,\boldsymbol{\rho}}}
\left[
V_{\boldsymbol{\rho}}(\vgamma)
\right],
\]
where
\[
P_{t,\boldsymbol{\rho}}(\vgamma)
=
\frac{
P(\vgamma)
e^{-tV_{\boldsymbol{\rho}}(\vgamma)}
}{
\E_P[
e^{-tV_{\boldsymbol{\rho}}(\vgamma)}
]
}
\]
defines a probability distribution over
$\{0,1\}^p$.
To bound the second derivative, note that
the model space $\{0,1\}^p$ is finite.
Moreover, Assumption~\ref{ass:PIP-convergence}(i)
guarantees that the posterior has full support,
while Assumption~\ref{ass:PIP-convergence}(ii)
ensures that all variational inclusion
probabilities are bounded away from zero
and one. Consequently, there exist constants
$0<u_-<u_+<\infty$ such that
\[
u_-
\leq U_{\boldsymbol{\rho}}(\vgamma)
\leq u_+,
\]
uniformly over $\vgamma\in\{0,1\}^p$
and $\boldsymbol{\rho}\in\mathcal R$.
Since
$\E_{q_{\boldsymbol{\rho}}}
[U_{\boldsymbol{\rho}}(\vgamma)]=1$,
these constants may be chosen such that
$u_-\leq 1\leq u_+$.
It follows that
\[
|V_{\boldsymbol{\rho}}(\vgamma)|
\leq B,
\qquad
B=\max\{|\log u_-|,|\log u_+|\}<\infty.
\]
Consequently,
\[
0\leq g_{\boldsymbol{\rho}}''(t)
\leq
\E_{P_{t,\boldsymbol{\rho}}}
\left[V_{\boldsymbol{\rho}}(\vgamma)^2\right]
\leq B^2,
\]
uniformly over $\boldsymbol{\rho}\in\mathcal R$
and $0\leq t\leq\alpha_0$.
The Taylor remainder therefore satisfies
\[
\sup_{\boldsymbol{\rho}\in\mathcal R}
|R_{\boldsymbol{\rho}}(\alpha)|
\leq
\frac{B^2}{2}\alpha^2,
\qquad
0<\alpha\leq\alpha_0.
\]

It remains to connect $K(\boldsymbol{\rho})$
to the variational R\'enyi bound and bound
the resulting error. From \eqref{eq:VR_IWAE_alt},
\[
\log Z-
\mathcal L_{\mathrm R}^{(\alpha)}
\left(\operatorname{logit}(\boldsymbol{\rho})\right)
=
-\frac{g_{\boldsymbol{\rho}}(\alpha)}
{1-\alpha},
\]
and hence
\[
\frac{1-\alpha}{\alpha}
\left(
\log Z-
\mathcal L_{\mathrm R}^{(\alpha)}
\left(\operatorname{logit}(\boldsymbol{\rho})\right)
\right)
=
K(\boldsymbol{\rho})
-
\frac{R_{\boldsymbol{\rho}}(\alpha)}{\alpha}.
\]
Thus,
\begin{align*}
\left|
\frac{1-\alpha}{\alpha}
\left(
\log Z-
\mathcal L_{\mathrm R}^{(\alpha)}
\left(\operatorname{logit}(\boldsymbol{\rho})\right)
\right)
-K(\boldsymbol{\rho})
\right|
&=
\frac{|R_{\boldsymbol{\rho}}(\alpha)|}{\alpha}
\leq \frac{B^2}{2}\alpha.
\end{align*}
Since the remainder bound holds uniformly over
$\boldsymbol{\rho}\in\mathcal R$, setting
$C_1=B^2/2$ and taking the supremum yields
\begin{equation}
\sup_{\boldsymbol{\rho}\in\mathcal R}
\left|
\frac{1-\alpha}{\alpha}
\left(
\log Z-
\mathcal L_{\mathrm R}^{(\alpha)}
\left(\operatorname{logit}(\boldsymbol{\rho})\right)
\right)
-K(\boldsymbol{\rho})
\right|
\leq C_1\alpha,
\label{eq:proof-renyi-KL-uniform}
\end{equation}
for all $0<\alpha\leq\alpha_0$.

\medskip
\noindent\textbf{(b) Uniform approximation error of
forward KL by a rescaled VR-IWAE gap.}
We now extend the uniform approximation of
forward KL established in part (a) to a
rescaled VR-IWAE gap by controlling the
additional finite-$N$ error between the
VR-IWAE bound and the variational
R\'enyi bound.
For fixed $\alpha$ and variational parameters,
\citet[Theorem~3]{daudel2023alpha}
establish an asymptotic $O(N^{-1})$ gap
between the VR-IWAE bound and the
variational R\'enyi bound, generalising
the corresponding IWAE result in
\citet[Theorem~3]{Domke2018}.
However, their asymptotic expansion does
not guarantee uniform control over
$\boldsymbol{\rho}\in\mathcal R$ and
$0<\alpha\leq\alpha_0$, as required to
establish Theorem~\ref{thm:PIP-convergence},
since the optimal variational probabilities
$\widehat{\boldsymbol{\rho}}_{N,\alpha}$
generally depend on both $N$ and $\alpha$. Instead, we establish a non-asymptotic bound
of the same order in $N$, uniformly over
$\boldsymbol{\rho}\in\mathcal R$ and
$0<\alpha\leq\alpha_0$, albeit under
the more restrictive assumptions of a
finite model space and variational inclusion
probabilities bounded away from zero and one.

For independent draws
$\vgamma_i\sim q_{\boldsymbol{\rho}}$,
$i=1,\ldots,N$, with
$U_{\boldsymbol{\rho}}(\cdot)$ defined in
\eqref{eq:norm_weights}, let
\[
X_i
=
U_{\boldsymbol{\rho}}(\vgamma_i)^{1-\alpha},
\qquad
\overline X_N
=
\frac{1}{N}\sum_{i=1}^N X_i,
\qquad
\mu_{\boldsymbol{\rho},\alpha}
=
\E_{q_{\boldsymbol{\rho}}}[X_i].
\]
Using $w_{\vlambda}(\vgamma;\mathbf y)
=ZU_{\boldsymbol{\rho}}(\vgamma)$,
the VR-IWAE bound in
\eqref{eq:VR-IWAE-bound} can be expressed as
\begin{align*}
\ell_N^{(\alpha)}
\left(\operatorname{logit}(\boldsymbol{\rho})\right)
&=
\log Z+
\frac{1}{1-\alpha}
\E\left[
\log\left(
\frac{1}{N}\sum_{i=1}^N
U_{\boldsymbol{\rho}}(\vgamma_i)^{1-\alpha}
\right)
\right]\\
&=
\log Z+
\frac{1}{1-\alpha}
\E\left[\log\overline X_N\right].
\end{align*}
Similarly, the variational R\'enyi bound in
\eqref{eq:renyi-bound} can be written as
\[
\mathcal L_{\mathrm R}^{(\alpha)}
\left(\operatorname{logit}(\boldsymbol{\rho})\right)
=
\log Z+
\frac{1}{1-\alpha}
\log\mu_{\boldsymbol{\rho},\alpha}.
\]
Thus, the difference between the variational R\'enyi
bound and the VR-IWAE bound is
\begin{equation}
\mathcal L_{\mathrm R}^{(\alpha)}
\left(\operatorname{logit}(\boldsymbol{\rho})\right)
-
\ell_N^{(\alpha)}
\left(\operatorname{logit}(\boldsymbol{\rho})\right)
=
\frac{1}{1-\alpha}
\left[
\log\mu_{\boldsymbol{\rho},\alpha}
-
\E\log\overline X_N
\right].
\label{eq:proof-VR-IWAE-gap}
\end{equation}
By Jensen's inequality, this difference is
nonnegative.

To bound the difference in \eqref{eq:proof-VR-IWAE-gap}, recall from part (a)
that $0<u_-\leq1\leq u_+<\infty$.
Since $0<1-\alpha<1$, we have
\[
u_-
\leq u_-^{1-\alpha}
\leq X_i
\leq u_+^{1-\alpha}
\leq u_+,
\]
uniformly over $\boldsymbol{\rho}\in\mathcal R$
and $0<\alpha\leq\alpha_0$. A second-order Taylor expansion of the logarithm
around $\mu_{\boldsymbol{\rho},\alpha}$,
with the remainder bounded using
$|(\log x)''|=1/x^2\leq 1/u_-^2$
for $x\geq u_-$, gives
\[
0\leq
\log\mu_{\boldsymbol{\rho},\alpha}
-
\E\log\overline X_N
\leq
\frac{\mathbb V(\overline X_N)}{2u_-^2}
=
\frac{\mathbb V(X_1)}{2Nu_-^2}.
\]
Substituting this bound into
\eqref{eq:proof-VR-IWAE-gap}, the difference
between the variational R\'enyi bound and
the VR-IWAE objective is therefore bounded
above by
$\mathbb V(X_1)/(2Nu_-^2(1-\alpha))$.
Since $\mathbb V(X_1)\leq u_+^2$
and $1-\alpha\geq1-\alpha_0$, setting
\[
C_2=
\frac{u_+^2}{2u_-^2(1-\alpha_0)},
\]
and taking the supremum over
$\boldsymbol{\rho}\in\mathcal R$, we obtain
\begin{equation}
0\leq
\sup_{\boldsymbol{\rho}\in\mathcal R}
\left|
\mathcal L_{\mathrm R}^{(\alpha)}
\left(\operatorname{logit}(\boldsymbol{\rho})\right)
-
\ell_N^{(\alpha)}
\left(\operatorname{logit}(\boldsymbol{\rho})\right)
\right|
\leq
\frac{C_2}{N}.
\label{eq:proof-VR-IWAE-uniform}
\end{equation}

To combine the two approximation errors, first write
\begin{align*}
\frac{1-\alpha}{\alpha}
\left(\log Z-\ell_N^{(\alpha)}(\operatorname{logit}(\boldsymbol{\rho}))\right)
-K(\boldsymbol{\rho})
&=
\left[
\frac{1-\alpha}{\alpha}
\left(\log Z-\mathcal L_{\mathrm R}^{(\alpha)}(\operatorname{logit}(\boldsymbol{\rho}))\right)
-K(\boldsymbol{\rho})
\right]\\
&\quad+
\frac{1-\alpha}{\alpha}
\left(
\mathcal L_{\mathrm R}^{(\alpha)}(\operatorname{logit}(\boldsymbol{\rho}))
-\ell_N^{(\alpha)}(\operatorname{logit}(\boldsymbol{\rho}))
\right).
\end{align*}
Applying the triangle inequality and taking
the supremum over $\boldsymbol{\rho}\in\mathcal R$,
\eqref{eq:proof-renyi-KL-uniform} and
\eqref{eq:proof-VR-IWAE-uniform} yield
\begin{align}
\sup_{\boldsymbol{\rho}\in\mathcal R}
\left|
\frac{1-\alpha}{\alpha}
\left[
\log Z-
\ell_N^{(\alpha)}
\left(\operatorname{logit}(\boldsymbol{\rho})\right)
\right]
-K(\boldsymbol{\rho})
\right|
&\leq
C_1\alpha+
\frac{(1-\alpha)C_2}{N\alpha}
\nonumber\\
&\leq
C_1\alpha+\frac{C_2}{N\alpha},
\label{eq:proof-VR-IWAE-forward-KL}
\end{align}
since $1-\alpha\leq1$. Thus,
\eqref{eq:proof-VR-IWAE-forward-KL}
establishes a non-asymptotic upper bound
on the approximation error between the
rescaled VR-IWAE gap and the forward KL
divergence $K(\boldsymbol{\rho})$.
The bound holds uniformly over
$\boldsymbol{\rho}\in\mathcal R$ and for
all $N\geq1$ and $0<\alpha\leq\alpha_0$,
with constants $C_1$ and $C_2$ independent
of $\boldsymbol{\rho}$, $N$ and $\alpha$.

\medskip
\noindent\textbf{Step 2: Bounding forward KL
and PIP errors under global optimisation.}

\noindent We first use the uniform approximation bound
established in Step~1 to control the difference
between the forward KL divergence evaluated
at the VR-IWAE optimiser
$\widehat{\boldsymbol{\rho}}_{N,\alpha}$
and the minimum forward KL divergence
$K(\boldsymbol{\rho}^\star)$ established
in the preliminaries.

Let
\[
F_{N,\alpha}(\boldsymbol{\rho})
=
\frac{1-\alpha}{\alpha}
\left[
\log Z-
\ell_N^{(\alpha)}
\left(\operatorname{logit}(\boldsymbol{\rho})\right)
\right]
\]
denote the rescaled VR-IWAE gap. Since $(1-\alpha)/\alpha>0$ and $\log Z$
does not depend on $\boldsymbol{\rho}$,
Assumption~\ref{ass:PIP-convergence}(iii)
implies that
$\widehat{\boldsymbol{\rho}}_{N,\alpha}$
minimises $F_{N,\alpha}$ over $\mathcal R$.
Consequently,
$F_{N,\alpha}(\widehat{\boldsymbol{\rho}}_{N,\alpha})
\leq F_{N,\alpha}(\boldsymbol{\rho}^\star)$.
Using this global optimality condition
and the uniform bound in
\eqref{eq:proof-VR-IWAE-forward-KL} twice,
we obtain
\begin{align}
K\left(\widehat{\boldsymbol{\rho}}_{N,\alpha}\right)
-K(\boldsymbol{\rho}^\star)
&=
\left[
K\left(\widehat{\boldsymbol{\rho}}_{N,\alpha}\right)
-F_{N,\alpha}\left(\widehat{\boldsymbol{\rho}}_{N,\alpha}\right)
\right]
\nonumber\\
&\quad+
\left[
F_{N,\alpha}\left(\widehat{\boldsymbol{\rho}}_{N,\alpha}\right)
-F_{N,\alpha}(\boldsymbol{\rho}^\star)
\right]
\nonumber\\
&\quad+
\left[
F_{N,\alpha}(\boldsymbol{\rho}^\star)
-K(\boldsymbol{\rho}^\star)
\right]
\nonumber\\
&\leq
\left|
K\left(\widehat{\boldsymbol{\rho}}_{N,\alpha}\right)
-F_{N,\alpha}\left(\widehat{\boldsymbol{\rho}}_{N,\alpha}\right)
\right|
+
\left|
F_{N,\alpha}(\boldsymbol{\rho}^\star)
-K(\boldsymbol{\rho}^\star)
\right|
\nonumber\\
&\leq
2C_1\alpha+\frac{2C_2}{N\alpha}.
\label{eq:proof-KL-excess}
\end{align}

Finally, we translate this bound on the forward
KL difference into a bound on the PIP approximation
errors. By \eqref{eq:proof-forward-KL-projection},
the difference between the forward KL
divergence at the VR-IWAE optimiser
$\widehat{\boldsymbol{\rho}}_{N,\alpha}$
and its minimum at $\boldsymbol{\rho}^\star$
is a sum of Bernoulli KL divergences.
Pinsker's inequality for Bernoulli distributions is
\[
\operatorname{KL}
\left(
\operatorname{Bernoulli}(u)
\|
\operatorname{Bernoulli}(v)
\right)
\geq
2\operatorname{TV}
\left(
\operatorname{Bernoulli}(u),
\operatorname{Bernoulli}(v)
\right)^2
=
2(u-v)^2.
\]
Applying this inequality together with
\eqref{eq:proof-forward-KL-projection}
and \eqref{eq:proof-KL-excess}, we obtain
\begin{align*}
2\sum_{j=1}^p
\left(
\widehat\rho_{j,N,\alpha}
-
\rho_j^\star
\right)^2
&\leq
\sum_{j=1}^p
\operatorname{KL}
\left(
\operatorname{Bernoulli}(\rho_j^\star)
\|
\operatorname{Bernoulli}(\widehat\rho_{j,N,\alpha})
\right)\\
&=
K\left(\widehat{\boldsymbol{\rho}}_{N,\alpha}\right)
-K(\boldsymbol{\rho}^\star)\\
&\leq
2C_1\alpha+\frac{2C_2}{N\alpha}.
\end{align*}
It follows that
\begin{align*}
\max_{1\leq j\leq p}
\left|
\widehat\rho_{j,N,\alpha}
-
\Pr(\gamma_j=1|\mathbf y)
\right|
&\leq
\sqrt{
C_1\alpha+\frac{C_2}{N\alpha}
}\\
&\leq
C\sqrt{
\alpha+\frac{1}{N\alpha}
},
\end{align*}
where $C=\sqrt{\max\{C_1,C_2\}}$
is independent of $N$ and $\alpha$.
This establishes \eqref{eq:optimal-PIP-bound}.
\end{proof}

\begin{proof}[Proof of Proposition~\ref{prop:collapsed-VR-IWAE}]

The corresponding joint importance weight under the joint variational distribution
$q_{\vlambda}^{\mathrm J}(\vgamma,\boldsymbol\psi)
=q_{\vlambda}(\vgamma)r(\boldsymbol\psi|\vgamma)$
is given by
\begin{equation}
w_{\vlambda}^{\mathrm J}(\vgamma,\boldsymbol\psi;\mathbf y)
=
\frac{
p(\mathbf y,\vgamma,\boldsymbol\psi)
}{
q_{\vlambda}(\vgamma)r(\boldsymbol\psi|\vgamma)
}.
\label{eq:joint_weights}
\end{equation}
Let $\vgamma_1,\ldots,\vgamma_N$ be independent draws from
$q_{\vlambda}$ and, conditional on $\vgamma_{1:N}$,
let $\boldsymbol\psi_i$ be independently drawn from
$r(\cdot|\vgamma_i)$, $i=1,\ldots,N$.
Applying Jensen's inequality yields
\begin{align}\label{eq:Jensen_psi_given_gamma}
&\E_{\boldsymbol\psi_{1:N}|\vgamma_{1:N}}
\left[
\frac{1}{1-\alpha}
\log\left(
\frac{1}{N}
\sum_{i=1}^N
w_{\vlambda}^{\mathrm J}
(\vgamma_i,\boldsymbol\psi_i;\mathbf y)^{1-\alpha}
\right)
\right]\nonumber\\
&\qquad\leq
\frac{1}{1-\alpha}
\log\left(
\frac{1}{N}
\sum_{i=1}^N
\left\{
\E_{r(\cdot|\vgamma_i)}
\left[
w_{\vlambda}^{\mathrm J}
(\vgamma_i,\boldsymbol\psi_i;\mathbf y)
\right]
\right\}^{1-\alpha}
\right).
\end{align}
The inequality is applicable since, for $0<\alpha<1$,
the function
\[
(x_1,\ldots,x_N)
\mapsto
\frac{1}{1-\alpha}
\log\left(
\frac{1}{N}
\sum_{i=1}^N x_i^{1-\alpha}
\right)
\]
is jointly concave, as $x^{1-\alpha}$ is concave
and the logarithm is concave and increasing.
This concavity extends to nonnegative arguments
by assigning the value $-\infty$ when all
arguments are zero.

By the support condition on
$r(\boldsymbol\psi|\vgamma)$, the conditional
expectation appearing on the right-hand side of
\eqref{eq:Jensen_psi_given_gamma} satisfies
\begin{align}
\E_{r(\cdot|\vgamma_i)}
\left[
w_{\vlambda}^{\mathrm J}
(\vgamma_i,\boldsymbol\psi_i;\mathbf y)
\right]
&=
\frac{1}{q_{\vlambda}(\vgamma_i)}
\int p(\mathbf y,\vgamma_i,\boldsymbol\psi_i)
\,d\boldsymbol\psi_i \nonumber\\
&=
\frac{p(\mathbf y,\vgamma_i)}
{q_{\vlambda}(\vgamma_i)}
=
w_{\vlambda}(\vgamma_i;\mathbf y).
\label{eq:cond_expectation_joint_weight}
\end{align}
Substituting this identity into
\eqref{eq:Jensen_psi_given_gamma}, taking expectations
over $\vgamma_{1:N}$ on both sides, and applying the
tower property on the left-hand side yields
\begin{equation}
\ell_{N,\mathrm J}^{(\alpha)}(\vlambda,r)
\leq
\ell_N^{(\alpha)}(\vlambda).
\label{eq:proof-collapsed-VR-IWAE}
\end{equation}
Furthermore, from \eqref{eq:VR-IWAE-ordering},
$\ell_N^{(\alpha)}(\vlambda)\leq\log p(\mathbf y)$.

Finally, applying Lemma~\ref{lem:VR-IWAE-PIP}
and the ordering in \eqref{eq:proof-collapsed-VR-IWAE}
yields
\[
\left|
\Pr(\gamma_j=1|\mathbf y)-\rho_j
\right|
\leq
\sqrt{
\frac{
\log p(\mathbf y)-\ell_N^{(\alpha)}(\vlambda)
}{
2\alpha
}
}
\leq
\sqrt{
\frac{
\log p(\mathbf y)
-\ell_{N,\mathrm J}^{(\alpha)}(\vlambda,r)
}{
2\alpha
}
},
\]
for every $j=1,\ldots,p$.
\end{proof}

\begin{proof}[Proof of Proposition~\ref{prop:collapsed-IS}]
As established in \eqref{eq:cond_expectation_joint_weight} in the proof of
Proposition~\ref{prop:collapsed-VR-IWAE}, the
collapsed importance weight satisfies
\[
\E_{r(\cdot|\vgamma)}
\left[
w_{\vlambda}^{\mathrm J}
(\vgamma,\boldsymbol\psi;\mathbf y)
\right]
=
w_{\vlambda}(\vgamma;\mathbf y).
\]
Combining this with the law of total variance,
\begin{align}
\mathbb V_{q_{\vlambda}^{\mathrm J}}
\left[w_{\vlambda}^{\mathrm J}(\vgamma,\boldsymbol\psi;\mathbf y)\right]
&=
\E_{q_{\vlambda}}
\left[
\mathbb V_{r(\cdot|\vgamma)}
\left[w_{\vlambda}^{\mathrm J}(\vgamma,\boldsymbol\psi;\mathbf y)\right]
\right]
+
\mathbb V_{q_{\vlambda}}
\left[
\E_{r(\cdot|\vgamma)}
\left[w_{\vlambda}^{\mathrm J}(\vgamma,\boldsymbol\psi;\mathbf y)\right]
\right]\nonumber\\
&=
\E_{q_{\vlambda}}
\left[
\mathbb V_{r(\cdot|\vgamma)}
\left[w_{\vlambda}^{\mathrm J}(\vgamma,\boldsymbol\psi;\mathbf y)\right]
\right]
+
\mathbb V_{q_{\vlambda}}
\left[w_{\vlambda}(\vgamma;\mathbf y)\right]\nonumber\\
&\geq
\mathbb V_{q_{\vlambda}}
\left[w_{\vlambda}(\vgamma;\mathbf y)\right].
\label{eq:variance_weights_inequality}
\end{align}
This establishes the ordering of the variances.

For the ordering of the effective sample sizes, the strong law of
large numbers gives, as $M\to\infty$,
\[
\frac{\operatorname{ESS}_M}{M}
=
\frac{
\left(M^{-1}\sum_{m=1}^M w_m\right)^2
}{
M^{-1}\sum_{m=1}^M w_m^2
}
\xrightarrow{\mathrm{a.s.}}
\frac{(\E[w])^2}{\E[w^2]},
\]
provided the importance weights have finite
second moments, which holds by assumption.
Moreover,
\[
\E_{q_{\vlambda}}
\left[w_{\vlambda}(\vgamma;\mathbf y)\right]
=
\E_{q_{\vlambda}^{\mathrm J}}
\left[w_{\vlambda}^{\mathrm J}(\vgamma,\boldsymbol\psi;\mathbf y)\right]
=
p(\mathbf y).
\]
Together with \eqref{eq:variance_weights_inequality},
this implies
\[
\E_{q_{\vlambda}}
\left[w_{\vlambda}(\vgamma;\mathbf y)^2\right]
\leq
\E_{q_{\vlambda}^{\mathrm J}}
\left[w_{\vlambda}^{\mathrm J}(\vgamma,\boldsymbol\psi;\mathbf y)^2\right].
\]
Consequently,
\[
\frac{p(\mathbf y)^2}
{\E_{q_{\vlambda}}
\left[w_{\vlambda}(\vgamma;\mathbf y)^2\right]}
\geq
\frac{p(\mathbf y)^2}
{\E_{q_{\vlambda}^{\mathrm J}}
\left[w_{\vlambda}^{\mathrm J}(\vgamma,\boldsymbol\psi;\mathbf y)^2\right]},
\]
which establishes
\eqref{eq:collapsed-ESS-inequality}
almost surely.
\end{proof}

\end{document}